\documentclass[aps,prl,
reprint,
groupedaddress]{revtex4-2}
\usepackage{graphicx}% Include figure files
\usepackage{dcolumn}% Align table columns on decimal point
\usepackage{booktabs}
\usepackage{tabularx}
\usepackage{bm}% bold math
\usepackage{color}
\usepackage{lipsum, babel}

\begin{document}

%\preprint{}

\title{Complementary experimental methods to obtain thermodynamic parameters of protein ligand systems}


\author{Shilpa Mohanakumar}
\affiliation{%
	IBI-4:Biomacromolecular Systems and Processes, Forschungszentrum Jülich GmbH, D-52428 Jülich, Germany \\
}
\author{Namkyu Lee}%
\affiliation{%
	IBI-4:Biomacromolecular Systems and Processes, Forschungszentrum Jülich GmbH, D-52428 Jülich, Germany \\
}
\author{Simone Wiegand}
\email{s.wiegand@fz-juelich.de}
\affiliation{%
	IBI-4:Biomacromolecular Systems and Processes, Forschungszentrum Jülich GmbH, D-52428 Jülich, Germany \\
}
\altaffiliation[Also at ]{Chemistry Department – Physical Chemistry, University Cologne, D-50939 Cologne, Germany.}%


\date{\today}

\begin{abstract}
In recent years, thermophoresis has emerged as a promising tool for quantifying biomolecular interactions. The underlying physical effect is still not understood. To gain deeper insight, we investigate whether non-equilibrium coefficients can be related to equilibrium properties. Therefore, we compare thermophoretic data measured by thermal diffusion forced Rayleigh scattering (TDFRS) (which is a non-equilibrium process) with thermodynamic data obtained by isothermal titration calorimetry (ITC) (which is an equilibrium process). As a reference system, we studied chelation reaction between ethylenediaminetetraacetic acid (EDTA) and calcium chloride (CaCl$_2$) to relate the thermophoretic behavior quantified by the Soret coefficient $S_{\mathrm T}$ to the Gibb's free energy $\Delta G$ determined in the ITC experiment using an expression proposed by Eastman [J. Am. Chem. Soc. {\bf{50}}, 283 (1928)]. Finally, we have studied the binding of the protein Bovine Carbonic Anhydrase I (BCA I) to two different benzenesulfonamide derivatives: 4-fluorobenzenesulfonamide (4FBS) and pentafluorobenzenesulfonamide (PFBS). For all three systems, we find that the Gibb' free energies calculated from $S_{\mathrm T}$ agree with $\Delta G$ from the ITC experiment. In addition, we also investigate the influence of fluorescent labeling, which allows measurements in a thermophoretic microfluidic cell. Re-examination of the fluorescently labeled system using ITC showed a strong influence of the dye on the binding behavior.
\end{abstract}

%\keywords{}

\maketitle

\section{Introduction}
Quantification of biomolecular interactions is extremely valuable in applications such as drug discovery and understanding molecular disease mechanisms. Several techniques have been developed providing binding affinities, kinetics and/or thermodynamics of the interactions \cite{Plach-2020-1}. One of the newer methods is MicroScale thermophoresis (MST) \cite{Jerabek-Willemsen-2014-101}. 
MST measures the thermophoretic movement of solutes in a temperature gradient by recording the fluorescent intensity. Typically, the binding constant is derived by using multiple capillaries with constant concentrations of protein and increasing ligand concentration. The capillaries are scanned consecutively, so that $K_{\mathrm{a}}$ can be determined, which gives access to the change in Gibb's free energy $\Delta G$. Since the technique uses fluorescent detection, either a fluorescent label is attached or the inherent fluorescence of the molecule of interest is detected \cite{Seidel-2012-10656}. The fluorescent labeling is very selective and allows low concentrations, but on the other hand the fluorescent label might influence the binding of the ligand. Although the underlying measurement effect is thermophoresis, the Soret and thermal diffusion coefficients are not determined.

Thermodiffusion is quantified by the Soret coefficient $S_{\mathrm T}=D_{\mathrm T}/D$, with the thermal diffusion coefficient $D_{\mathrm T}$ and the diffusion coefficient $D$ \cite{Kohler-2016-151,Niether-2019-503003}. A negative $S_{\mathrm T}$ indicates thermophilic behavior which means the solute accumulates on the warmer side. While $S_{\mathrm T}$ being positive (thermophobic) indicates a movement of the solute towards the colder side. Studies of aqueous systems suggest that the change in the thermodiffusive behavior is often connected with a variation in the hydration shells \cite{Niether-2016-4272,Niether-2018-044506,Niether-2018-1012}.
For certain solutions, a sign change from thermophilic to thermophobic behavior can be observed at a transition temperature $T^{*}$ \cite{Kishikawa-2010-740}. 
An empirical equation for diluted aqueous solutions proposed by Iacopini and Piazza \cite{Iacopini-2006-59} describes the temperature dependence by,
\begin{equation}
	\label{eq_Piazza}
	{S_{\rm{T}}}\left( T \right) = S_{\rm{T}}^\infty \left[ {1 - \exp \left( {\frac{{{T^ \ast } - T}}{{{T_0}}}} \right)} \right]\; ,
\end{equation}
where $ S_{{T}}^\infty$ is a constant value approached at high temperatures, $T^{\ast}$ is the temperature at which sign change of $S_{\mathrm T}$ occurs and $T_{0}$ indicates the curvature. Eq.\ref{eq_Piazza} describes how $S_{\mathrm T}$ increases with increasing temperature : $S_{\mathrm T}$ is low at lower temperatures to approach a plateau value at high temperatures \cite{Niether-2016-4272,Niether-2018-044506,Niether-2018-1012}. Solute-solvent interactions play a crucial role in the temperature sensitivity of $S_{\mathrm T}$. In aqueous solutions this contribution decreases with rising temperature due to breaking of hydrogen bonds \cite{Niether-2018-020001}.
%what have been studied in thermophoresis
For a number of aqueous systems, the difference of $S_{\mathrm T}$ at two different temperatures $\Delta S_{\mathrm T}$ has found to correlate with log $P$ (partition coefficient) \cite{Niether-2017-8483,Niether-2018-044506}. This indicates that the hydrophilicity of solute plays a crucial role in the thermophoretic behavior of aqueous systems. Log $P$ or the partition coefficient describes the concentration distribution of a solute between an aqueous and a 1-octanol phase in equilibrium. Thus $P$ is defined as 
\begin{equation}
	\label{parition_coefficient}
	P =\frac{[{\mathrm{solute}}]_{\mathrm{oil}}}{[{\mathrm{solute}}]_{\mathrm{water}}}
\end{equation}
Solutes which are highly hydrophilic (low or negative log $P$)  show a stronger change of $S_{\mathrm T}$ with temperature \cite{Niether-2018-044506}. At low temperatures hydrophilic solutes form many hydrogen bonds with water, while their number and strength decrease with increasing temperature. This means that at lower temperature there is a greater change in the hydration layer, which affects the Soret coefficient to a greater extent \cite{Schimpf-1989-1317,Piazza-2004-1616,Naumann-2013-5614}.

%protein -ligand systems

To investigate the thermophoretic behavior quantitatively we use Thermal Diffusion Forced Rayleigh Scattering (TDFRS). This is an optical method which analyses the diffraction efficiency of a refractive index grating due to temperature and concentration modulation. Ideally, the method is applied to binary mixtures, so biological systems with several components (buffer compounds) to stabilize the solution are more challenging because all compounds contribute to the refractive index contrast and complicate the analysis. So far only the strongly binding protein-ligand system streptavidin with biotin has been studied by TDFRS  \cite{Niether-2018-020001,Niether-2020-376}. The measurements were supported by neutron scattering experiments and also isothermal calorimetry data were included in the analysis  \cite{Sarter-2020-324,Sarter-2022-submitted}. Experiments showed
that the temperature sensitivity of the Soret coefficient was reduced for the complex compared to the free protein indicating that the complex was less hydrophilic leading to a larger entropy of the hydration layer. The outcome was in agreement with the neutron scattering data. The study of this particular system illustrates that thermodiffusion and its temperature dependence are highly sensitive to changes in the hydration layer. Although the exact mechanism of these changes cannot be evaluated by the study of a single system, measurements of similar systems can give us a more explicit picture on the conformational and hydration changes that occur upon ligand binding.

%What is ITC and what are the parameters that can be obtained\\
Isothermal titration calorimetry (ITC) is a standard method for any chemical (binding) reaction \cite{Baranauskiene-2009-2752}. It directly measures the heat released or consumed in the course of a molecular binding event. Besides thermodynamic parameters such as enthalpy $\Delta H$, entropy $\Delta S$, and Gibb's free enthalpie $\Delta G$ change, the equilibrium-binding affinity $K_{\mathrm{a}}$ and interaction stoichiometry can be determined. Among the biophysical characterization methods ITC offers the highest information content \cite{Plach-2020-1}.

A recently developed thermophoretic microfluidic cell was so far only tested with fluorescently labeled colloidal particles \cite{Lee-2022-123002}. In principle, the cell can also be used to monitor quantitatively the thermophoretic properties of fluorescently labeled free proteins and complexes as used in MST.

Although the thermophoretic behavior of the free  protein compared to the protein-ligand complex differs, the microscopic mechanism for this change is not yet understood. The underlying physical effect is one of the interesting unsolved puzzles in physical chemistry. Binding reactions are quite complex, strongly influenced by several factors like temperature, concentration, pH, ionic strength etc. and in turn influence the thermophoretic motion \cite{Wurger-2010-126601,Kohler-2016-151,Niether-2019-503003}. 
%Motivation of this work and theoretical background
In this work due to the complexity of the system and the physical effect, we study chemical binding reactions with TDFRS and ITC. Based on the results of the complementary methods we want to establish a relation between thermodynamic parameters obtained by ITC and thermophoretic properties measured with TDFRS. Additionally, selective ITC measurements and studies in a thermophoretic microfluidic cell were performed to investigate the influence of a fluorescence label on the binding and thermophoretic behavior.

To connect the thermodynamic parameter determined with ITC with the non-equilibrium coefficient derived from TDFRS experiments, we start from an early work by Eastman \cite{Eastman-1928-283}. In modern notation his approach connects the Soret coefficient, $S_{\mathrm T}$, to the Gibb's free energy as follows \cite{Eastman-1928-283,Wurger-2013-438}:
\begin{equation}
	\label{eq:eastman}
	S_{T} =\frac{1}{k_{\mathrm{B}}T} \frac{dG}{dT}
\end{equation}
This approach is not viewed uncritically, already de Groot wrote \cite{Groot-1966-book}, that Eastman's theory is "{\emph{... certainly not rigorous at all}}".  Integrating Eq.\ref{eq:eastman} with respect to temperature will give us access to a relation between $S_{\mathrm T}$ and $\Delta G$ for the individual compounds of the system (free protein, free ligand and complex). A detailed derivation can be found in the Supporting Information Sec.~1.
%( Sec.\ref{Sec:Mathematical_formulation}).
\begin{figure}[h!]
	\centering
	\includegraphics[width=8cm]{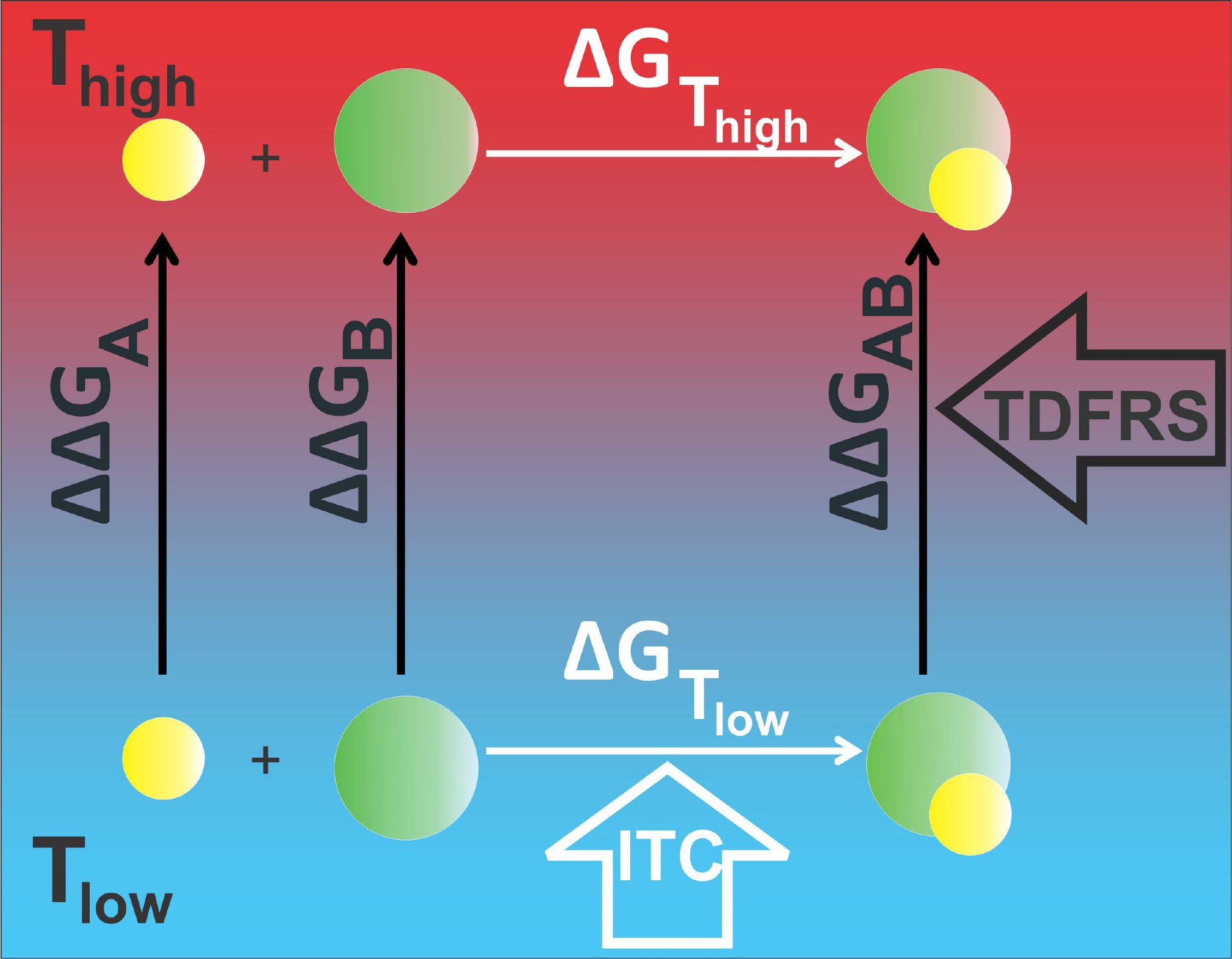}
	\caption{\label{fig:ITC+TDFRS_dG_calculation} Schematic illustration of the calculation of $\Delta G$ and $\Delta \Delta G$ from ITC and TDFRS respectively.}
\end{figure} 
How these individual contributions can be used to establish a relation between ITC and TDFRS measurements is illustrated in Fig.~\ref{fig:ITC+TDFRS_dG_calculation}. "A" and "B" correspond to the molecules which are used to form the complex "AB". We measure the free energy change $\Delta G$ at two different temperatures with ITC ($\Delta G_{T_{\mathrm{low}}}$ and $\Delta G_{T_{\mathrm{high}}}$). We hypothesize that $\Delta G_{T_{\mathrm{high}}}$ can be calculated from the free energy change at low temperature $\Delta G_{T_{\mathrm{low}}}$ measured by ITC and the differences in $\Delta\Delta G$ corresponding to two temperatures for the individual components probed by TDFRS using the following equation
\begin{equation}
	\label{eq:dG-T-high}
	\Delta G_{T_{\mathrm{high}}}= \Delta G_{T_{\mathrm{low}}}+ \Delta \Delta G_{\mathrm{AB}}-\Delta \Delta G_{\mathrm{A}}-\Delta \Delta G_{\mathrm{B}}.
\end{equation}
%EDTA+CaCl2-reference system
To test our hypothesis we use EDTA and CaCl$_2$ as reference system.
The chelation reaction between ethylenediaminetetraacetic acid (EDTA) and calcium chloride (CaCl$_2$) is a well known reaction which is used as a validation standard for ITC measurements \cite{Rafols-2016-354}. 
%protein-ligand system
In the next step we use the same formalism for the protein Bovine Carbonic Anhydrase I (BCA I) with two different ligands. The enzyme BCA I is responsible for the conversion of carbon dioxide to bicarbonate \cite{Maren-1987-255} and inhibitors of this enzyme are used for the treatment of glaucoma and epilepsy \cite{Cecchi-2005-5192}. Arylsulfonamides have the highest affinity and are mainly used as inhibitors for BCA I \cite{Innocenti-2008-1583,Supuran-2003-146b}. In our study we used  4-fluorobenzenesulfonamide (4FBS) and Pentafluorobenzenesulfonamide (PFBS) (cf. Fig.~\ref{fig:ligand_structure}). A previous study of BCA II, which is a variant of our enzyme, shows that PFBS binds approximately 25 times stronger than 4FBS at 25$^\circ$C \cite{Krishnamurthy-2007-94,Krishnamurthy-2008-946}. Therefore, we assume that the binding for these two ligands differs for BCA I as well, so that we can test our method for varying binding constants.

\begin{figure}[h!]
	\centering
	\includegraphics[width=7.4cm]{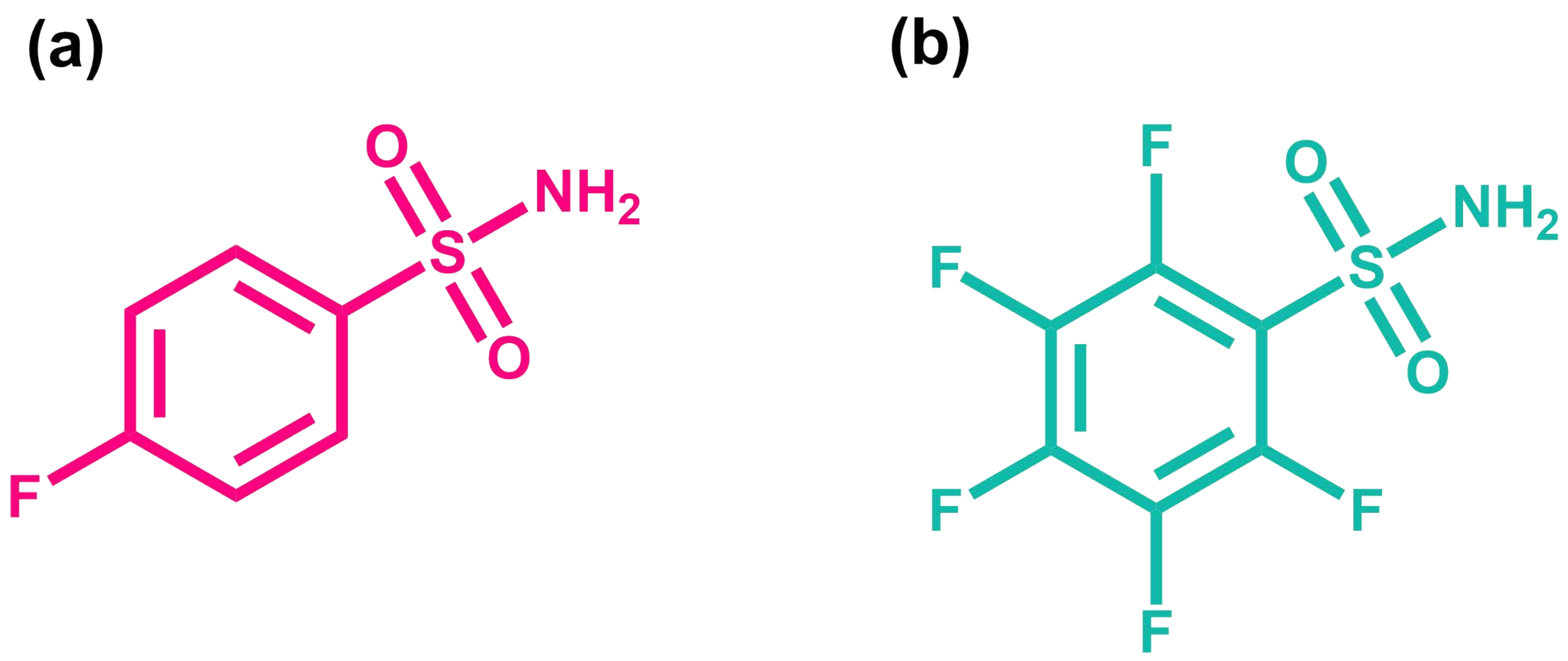}
	\caption{\label{fig:ligand_structure} Chemical structure of the two investigated ligands: (a) 4-fluorobenzenesulfonamide (4FBS) and (b) Pentafluorobenzenesulfonamide (PFBS).}
\end{figure}

%%%%%%%%%%%%%%%%%%%%%%%%%%%%%%%%%%%%%%%%%%
\section{Results and Discussion}
\subsection{EDTA-CaCl$_2$ system}
\subsubsection{TDFRS measurements}
\begin{figure}[h!]
	\centering
	\includegraphics[width=9 cm]{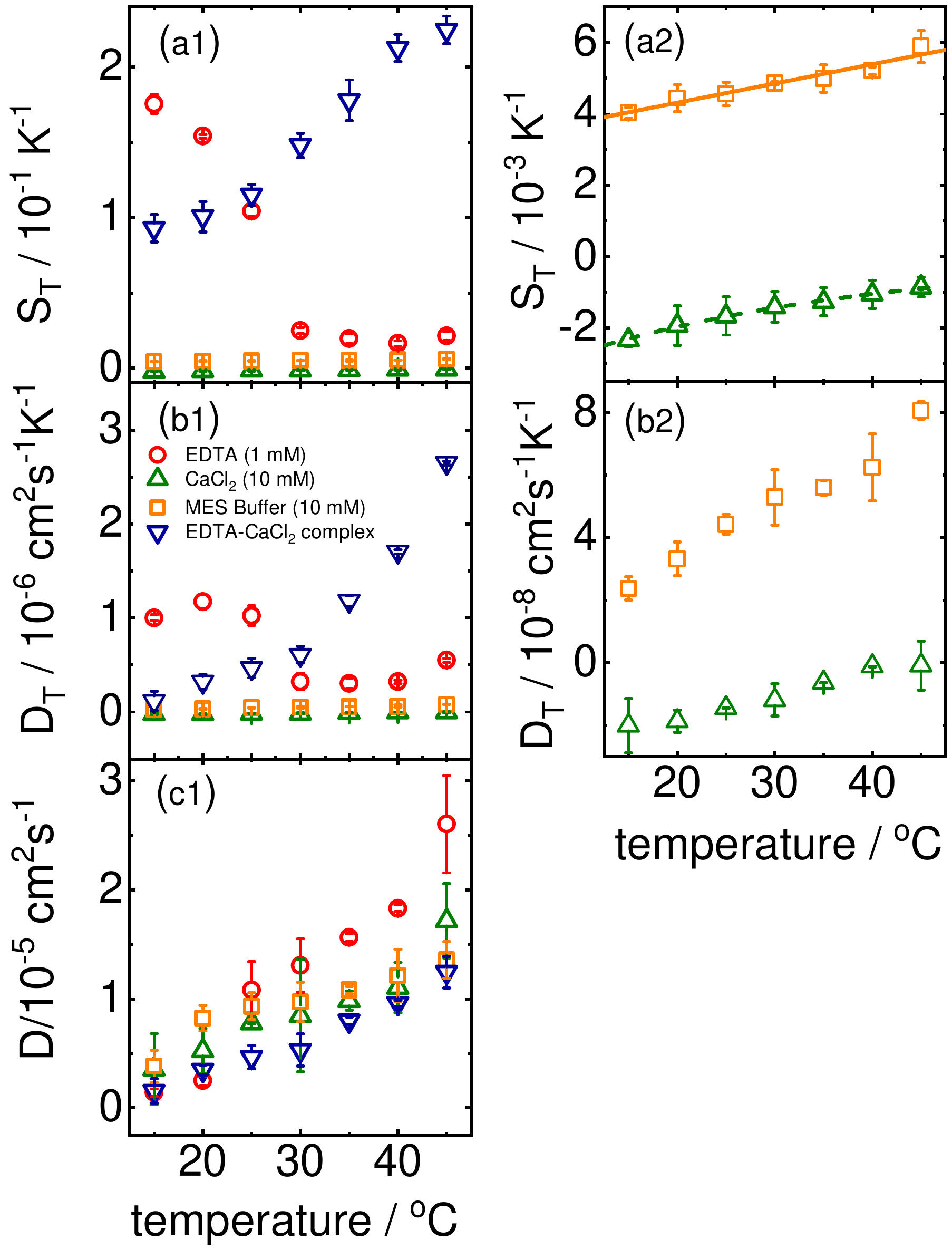}
	\caption{\label{fig:ST_vs_T_edta__calc2} Temperature dependence of (a1) $S_{\mathrm{T}}$, (b1) $D_{\mathrm{T}}$ and (c1) $D$ for EDTA (1mM), CaCl$_2$ (10mM), MES buffer (10mM) and EDTA-CaCl$_2$  complex. The error bars correspond to the standard deviation of the mean of repeated measurements. Figures on the right side is a zoomed in image of temperature dependence of (a2) $S_{\mathrm{T}}$ and (b2) $D_{\mathrm{T}}$ for CaCl$_2$ and MES buffer. Lines in (a2) corresponds to the fit according to Eq.\ref{eq_Piazza}.  }
\end{figure}

We conducted IR-TDFRS measurements for the individual components  EDTA, CaCl$_2$, MES Buffer and EDTA-CaCl$_2$ complex. Figure \ref{fig:ST_vs_T_edta__calc2} shows the temperature dependence of $S_{\mathrm{T}}$, $D_{\mathrm{T}}$ and $D$ for EDTA (1 mM), CaCl$_2$ (10 mM), MES buffer (10 mM) and EDTA-CaCl$_2$ complex.  

$S_{\mathrm{T}}$ of MES buffer is positive, while CaCl$_2$ in buffer displays thermophilic behavior ($S_{\mathrm{T}}<0$). For both systems the temperature dependence of $S_{\mathrm{T}}$ can be described by Eq.\ref{eq_Piazza}. The Soret coefficient of MES buffer and CaCl$_2$ in buffer is of the order of $10^{-3}$ K$^{-1}$, while $S_{\mathrm{T}}$ of EDTA and the complex EDTA-CaCl$_2$ are two orders of magnitude larger (cf. Fig.~\ref{fig:ST_vs_T_edta__calc2}(a1)). Therefore, we treat the solutions of EDTA and the complex (EDTA-CaCl$_2$) as quasi binary system analyzing the TDFRS data. The Soret coefficient of the complex shows an increase in $S_{\mathrm T}$ with temperature, but cannot be described by Eq.~\ref{eq_Piazza} as it has a turning point. $S_{\mathrm T}$ of EDTA decays with increasing temperature with an unusual pronounced drop between 25$^\circ$C and 30$^\circ$C. In the literature \cite{Sugiyama-2014-28,Minkevich-2006-1205,Yilmaz-2011-825,Pidard-1986-604}, there are works reporting a change of behavior in properties of several systems in presence of EDTA around 30$^\circ$C compared to that of room temperature, but so far no explanation has been developed.
A similar sudden change of $S_{\mathrm T}$ with temperature in the same temperature range has been reported for poly($N$-isoproplacrylamide) (PNiPAM) in water \cite{Kita-2005-4554}. PNiPAM is a temperature sensitive polymer showing a coil globule transition between 25$^\circ$C and 33$^\circ$C \cite{Qiu-2013-56,Bischofberger-2014-4377}. A small part of the drop of $S_{\mathrm{T}}$ is related to the increase of the diffusion coefficient, but the larger part is caused by the abrupt drop of $D_{\mathrm{T}}$ when the polymer coil collapses \cite{Kita-2005-4554}. In the case of of EDTA $D_{\mathrm{T}}$ also drops, while the diffusion coefficient of the small EDTA molecule does not show any unusual behavior and increases monotonously with temperature. The mechanism leading to the abrupt change in $D_{\mathrm{T}}$ of EDTA in water might have the same origin as in the case of PNiPAM as it happens in the same temperature range so that it is very likely influenced by hydrogen bonds. Bischofsberger {\em {et al.}} \cite{Bischofberger-2014-4377} argue that at higher temperatures the system minimizes its free energy by gaining entropy due releasing water molecules from the hydration shell. Although the microscopic mechanism is still unclear, this is further evidence that changes in water structure affect thermophoretic motion.

\subsubsection{ITC measurements}
As mentioned before, EDTA-CaCl$_2$ is a system that has been well studied and characterized using ITC at room temperature \cite{Rafols-2016-354,Griko-1999-79,Christensen-2003-7357}. For our goal we need binding parameters of the system in a wide temperature range.  Our results are summarized in Table \ref{table:binding_edta_cacl2}.
\\
\begin{table*}[h!] 
	\caption{Thermodynamic parameters of the binding reaction between EDTA and CaCl$_2$ measured using ITC at different temperatures by setting $m=1$ for the fit \label{table:binding_edta_cacl2}}
	\newcolumntype{C}{>{\centering\arraybackslash}X}
	\begin{tabularx}
		{\textwidth}{CCCC}
		\toprule
		Temperature ($^\circ$C)	& K$_d$ (nM)	& $\Delta$H (kJ/mol) \\
		\midrule
		20		& 510 $\pm$49		& -17 $\pm$0.3\\
		25		& 623$\pm$70.3			& -17.2$\pm$0.8\\
		30		& 699$\pm$55.5				& -17.3$\pm$0.7\\
		35 	& 852	$\pm$78.9		& -17.6$\pm$0.5 \\
		40		& 1210	$\pm$123		& -17.8 $\pm$0.5\\
		45		& 1570	$\pm$134		& -18 $\pm$0.8\\
		\bottomrule
	\end{tabularx}
\end{table*}

The reaction is found to be temperature sensitive and is more favored at lower temperatures. This is similar to what has been observed by Arena {\em {et al.}}, monitoring the association constant of the exothermic reaction between EDTA and Ca$^{2+}$ \cite{Arena-1983-3129}, where the found a decrease with increasing temperature.

\subsection{Protein-ligand system}

\subsubsection{TDFRS measurements}
\label{Sec:TDFRS_results}

\begin{figure}[h!]
	\centering
	\includegraphics[width=8 cm]{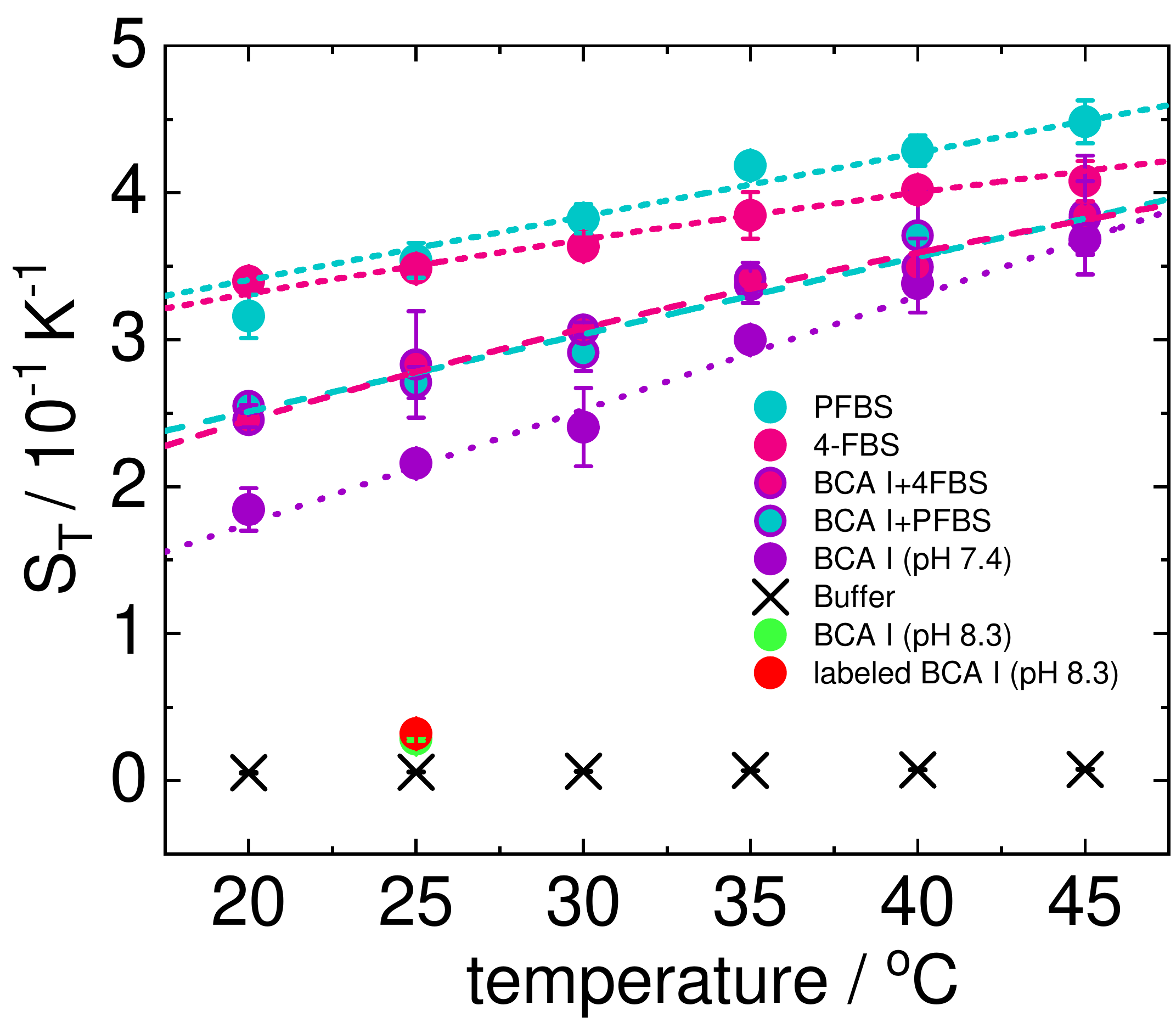}
	\caption{\label{fig:ST_vs_T_BCA_ligand} Temperature dependence of $S_{\mathrm T}$ for BCA I (10 $\mu$M, pH 7.4, violet bullets), BCA I (10 $\mu$M, pH 8.3, green bullets), labeled BCA I (pH 8.3, red bullets), 4FBS (110 $\mu$M, pink bullets)  and PFBS (110 $\mu$M, turquoise bullets), corresponding protein-ligand complex, sodium phosphate buffer (20 mM, Black cross). }
\end{figure}

Temperature dependence of the thermophoretic behavior of the free protein (BCA I), free ligands (4FBS and PFBS) and protein-ligand complexes is shown in Fig. \ref{fig:ST_vs_T_BCA_ligand}. As expected,  the Soret coefficient $S_{\mathrm T}$ of free BCA I changes significantly once the ligand binds. $S_{\mathrm T}$ of the complex is higher compared to that of the free protein. Increase in $S_{\mathrm T}$ with temperature of BCA I-ligand complex compared to free BCA I is different from that observed for the Streptavidin-biotin(STV-B) system \cite{Niether-2020-376}. For STV-B the difference between $S_{\mathrm T}$ of the free protein and complex increases with increasing temperature.  This was attributed to the stiffness of the protein at low temperatures so that the binding of the ligand (biotin) has a weaker effect at these temperatures \cite{Niether-2020-376}. In contrast to this, for both protein-ligand systems that we have studied the difference between $S_{\mathrm T}$ of free protein and complex decreases with temperature, so that it is almost negligible at high temperatures. This is an indication that the binding of both the ligands should become weaker with increasing temperature. This is in line with ITC measurements, which will be discussed in detail in Sec.\ref{Sec:protein_ligand_ITC}.

Hydrogen bonds  have a clear influence on the variation of  $S_{\mathrm T}$ with temperature. Change in  $S_{\mathrm T}$ with temperature is more evident, if the solute can form more hydrogen bonds with water \cite{Niether-2020-376,Niether-2019-503003}, therefore we conclude that the free protein is more hydrophilic than the protein-ligand complex (cf.~Fig.~\ref{fig:ST_vs_T_BCA_ligand}). So far, the temperature dependence of the thermophoretic behavior has only been studied for two other systems; STV-B and various unmethylated cyclodextrins with acetylsalicylic acid \cite{Niether-2020-376,Niether-2017-8483}. In both cases, the stronger temperature dependence of the free protein or host molecule indicates a lower hydrophilicity of the formed complexes. 

As we couldn't find studies which looked into the reaction mechanism of BCA I with the selected sulfonamide ligands, we compared Human Carbonic Anhydrase I (HCA I) with 4FBS and Bovine Carbonic Anhydrase II (BCA II) with both ligands which have been well characterized \cite{Dugad-1988-4310,Krishnamurthy-2007-94,Olander-1973-1616,Kernohan-1966-402,Olander-1970-5758}. The active site of the different variants of carbonic anhydrase protein (HCA I, BCA II) is the Zn$^{2+}$ ion that is tetrahedrally coordinated by three histidyl residues and a water molecule \cite{Vedani-1989-4075, Saito-2004-792}, to which sulfonamide ligands usually bind \cite{Abbate-2002-3583, Supuran-2003-146b}. In the literature, two scenarios of binding of sulfonamide ligands are discussed. The first suggests that sulfonamides are present in  the anionic form in their complexes with carbonic anhydrase \cite{Krishnamurthy-2007-94,Olander-1973-1616,Kernohan-1966-402,Olander-1970-5758,Lindskog-1968-453}, while the latter proposes neutral sulfonamides are bound to the active zinc ion \cite{Olander-1973-1616}. Detailed mechanism in both the cases has been discussed by Krishnamurthy {\em {et al.}}, \cite{Krishnamurthy-2007-94}. It has to be noted that in both possible scenarios a  water molecule is being released upon ligand binding. This implies that the complex is less hydrophilic than that of the free protein, which is what has also been concluded from the thermophoretic data.

In the literature, it has been reported that an increase in fluorination decreases the strength of hydrogen bond network between SO$_2$NH group and the active site of the target protein \cite{Krishnamurthy-2007-94}. This implies that the complex of BCA I with PFBS (which is highly fluorinated) should show a weaker temperature sensitivity of $S_{\mathrm T}$ compared to 4FBS. This is what we observe from our TDFRS measurements as we find; $\Delta S_{\mathrm T}$ = $S_{\mathrm T}$(45$^\circ$C)-$S_{\mathrm T}$(20$^\circ$C), $\Delta S_{\mathrm T}$(BCAI-4FBS)=0.139 K$^{-1}$ and $\Delta S_{\mathrm T}$(BCAI-PFBS)=0.128 K$^{-1}$.

\subsubsection{Isothermal titration calorimetry measurements}
\label{Sec:protein_ligand_ITC}

Thermodynamic parameters that have been obtained for the respective binding mechanisms at 25$^{\circ}$C are reported in Table~\ref{thermodynamic_parameters}. Fig.~\ref{fig:Kd_vs_T} shows an increase of the dissociation constants for both complexation reactions with temperature which supports the TDFRS measurements. Both ligands show a stoichiometry of 1:1 binding to the protein.

\begin{table}[h!] 
	\caption{Thermodynamic parameters of the binding reactions measured using ITC at 25$^{\circ}$C\label{thermodynamic_parameters}}
	\newcolumntype{C}{>{\centering\arraybackslash}X}
	\begin{tabularx}{\textwidth}{CCCC}
		\toprule
		System	& K$_d$ (nM)	& $\Delta$H (kJ/mol) & $\Delta$G (kJ/mol)\\
		\midrule
		BCA I + PFBS		& 127	$\pm$ 47.2		& -12.5 $\pm$ 0.8 & -37.4 $\pm$  2.8\\
		BCA I+ 4FBS	& 325 $\pm$ 58.7			& -32.7 $\pm$  0.4 & -37.5 $\pm$ 1.3\\
		\bottomrule
	\end{tabularx}
\end{table}

\begin{figure}[h!]
	\centering
	\includegraphics[width=7 cm]{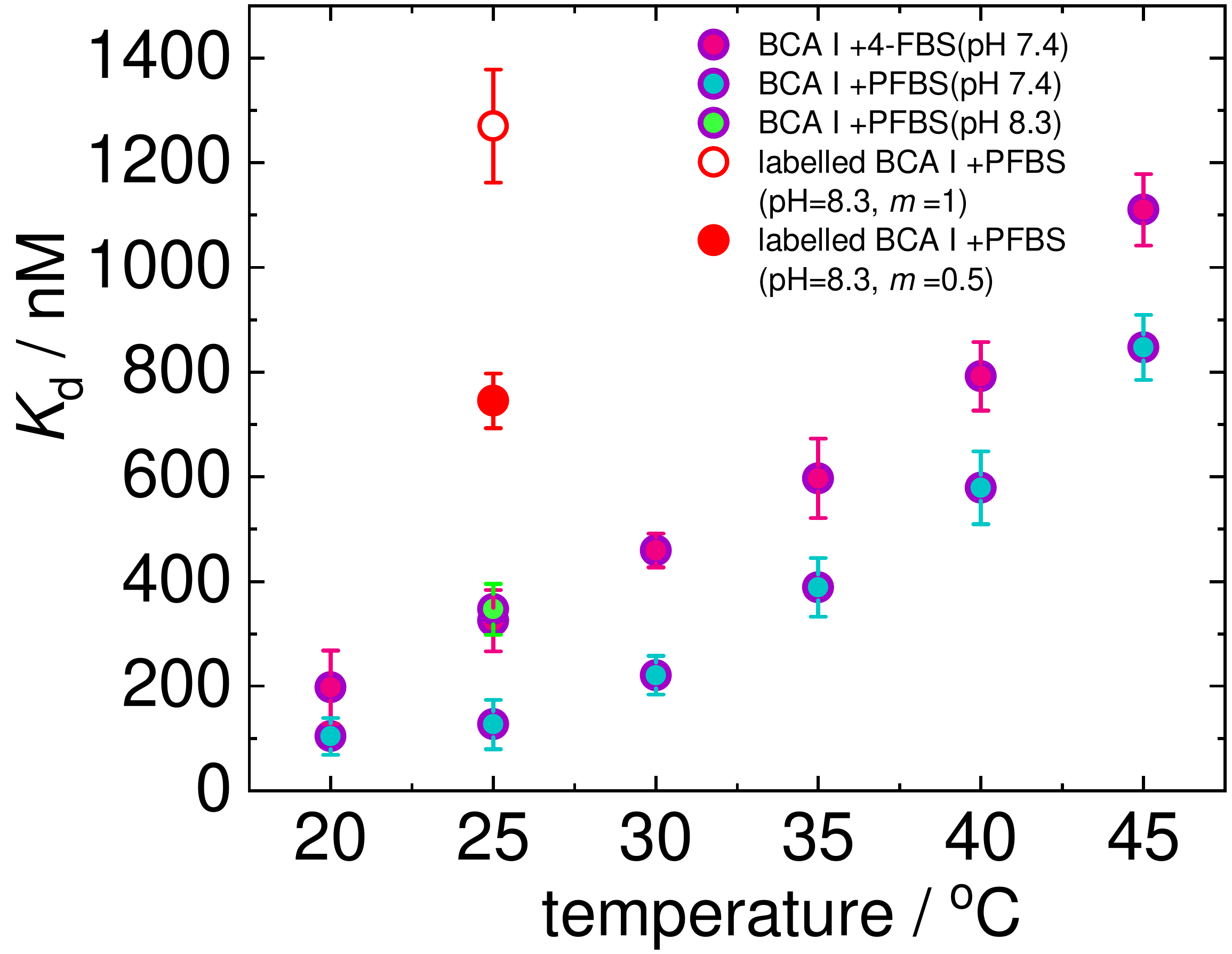}
	\caption{\label{fig:Kd_vs_T} Temperature dependence of K$_d$ for BCA I-4FBS and BCA I-PFBS complexes measured with ITC at pH=7.4. For comparison we show also a single measurement at 25$^\circ$C of the labeled and unlabeled BCA I-PFBS complex at pH=8.3. For the labeled BCA I-PFBS complex, we report two K$_d$
		values; red open circle (value that is obtained with $m=1$) and red closed circle (value that
		is obtained with $m=0.5$). More details about the difference in K$_d$ and stoichiometry values of two fits for labeled BCA I are discussed in Sec.\ref{Sec:microfluidic_cell}. }
\end{figure}
\noindent Increase in fluorine substitution is found to enhance the inhibitor power of sulfonamide ligands \cite{Hansch-1985-493}, implying that the more fluorine substituted ligand (PFBS) exhibits a higher association with BCA I, which is reflected by a lower K$_\mathrm{d}$ value, compared to 4FBS for all temperatures. Note that the dissociation constants of two ligands differ for BCA I only by a factor of 2.5, while for BCA II a factor of 25 has been reported \cite{Krishnamurthy-2007-94, Krishnamurthy-2008-946}.												

\subsubsection{Measurements with a thermophoretic microfluidic cell}
\label{Sec:microfluidic_cell}

We also used a thermophoretic microfluidic cell for measuring Soret coefficients \cite{Lee-2022-123002}. This requires the system to be fluorescent labeled to determine the concentration profile. The labeling process is explained in detail in the Supporting Information (cf. Sec.~3). Since the dye binds at a slightly higher pH=8.3, we performed additional TDFRS measurements at this pH with the labeled and unlabeled protein. We found  $S_{\mathrm{T}}=0.028\pm0.001$ K$^{-1}$ and $S_{\mathrm{T}}=0.032\pm0.001$ K$^{-1}$ for the unlabeled and labeled protein, respectively. Note, that the Soret coefficients measured at pH=8.3 are roughly an order of magnitude smaller than at pH=7.4 ($S_{\mathrm{T}}=0.216\pm0.003$ K$^{-1}$). The reason might be that with increasing pH the solute gets more negatively charged and can form more hydrogen bonds, which often leads to lower $S_{\mathrm{T}}$-values \cite{Niether-2019-503003}. The Soret coefficient $S_{\mathrm{T}}=0.018 $K$^{-1}$ measured in the microfluidic cell is roughly 40\% lower than the TDFRS-value and has a high uncertainty. The measured fluorescence intensity is at the detection limit due to  the low fraction of labeled proteins and decays due to photo bleaching. From repeated successful measurements we determine an uncertainty of 0.003 K$^{-1}$, but the real error might be higher due to systematic errors caused by bleaching.

\begin{figure}[h!]
	\centering
	\includegraphics[width=7 cm]{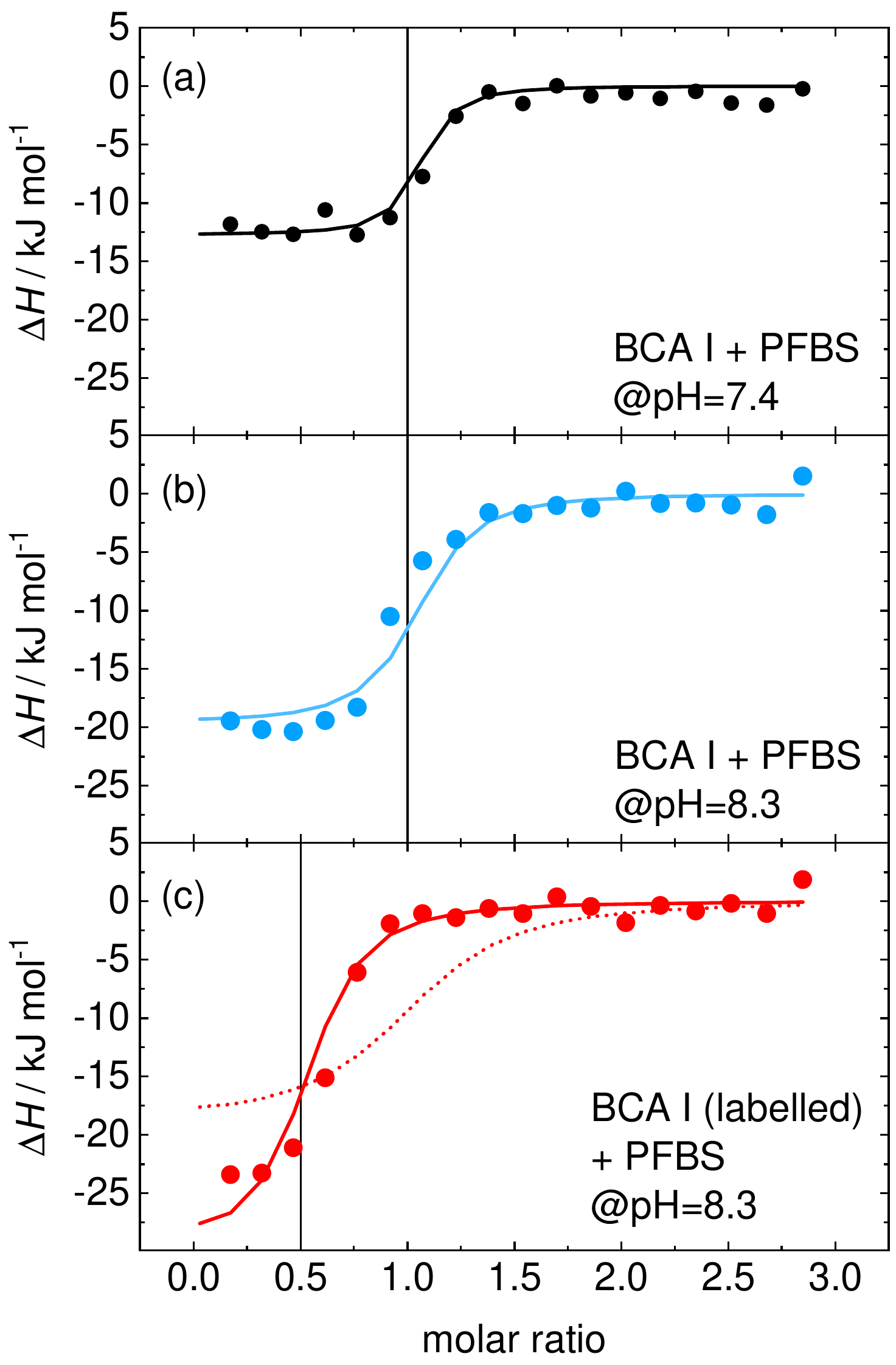}
	\caption{\label{fig:MasterplotITC} Molar change in enthalpy versus mole ratio of ligand over protein. (a) BCA I + PFBS at pH=7.4 (b) BCA I + PFBS at pH=8.3 and (c) the fluorescently labelled BCA I + PFBS at pH=8.3. Dotted and solid lines corresponds to fit with $m=1$ and $m=0.5$ respectively. All measurements have been performed at 25$^\circ$C.}
\end{figure}

To check the influence of the fluorescent label on the binding constant, we performed also ITC measurements. Since a change in pH is reported to affect the inhibitory power and activity of sulfonamides and protein, changes in the binding parameters are expected (cf.~Fig.~\ref{fig:MasterplotITC}) \cite{Henry-1943-175,Henry-1991-119}. An increase in pH, shows a decrease in association of PFBS with BCA I (cf.~Fig.~\ref{fig:Kd_vs_T}). Baronas {\em {et al.}} \cite{Baronas-2021-993} report a weak increase in the dissociation constant of carbonic anhydrase with primary sulfonamides when the pH changes from 7.4 to 8.3. Once the protein is labeled, the association is only 30\% compared to that of the unlabeled free protein at pH 8.3, so that we assume the dye blocks the binding site of the ligand (cf.~Fig.~\ref{fig:Kd_vs_T}). Additionally, the stoichiometry of ligand:protein changes from 1:1 to 1:2. A hypothesis for this behavior could be the existence of protein dimer, thus a single ligand binding to two proteins as it has been previously reported for lysozyme \cite{Darby-2017-1201}. Further experiments, e.g. using fluorescent correlation spectroscopy would have to be conducted to support this hypothesis.

In conclusion we refrained from more systematic measurements of fluorecently labeled proteins due to the change of the binding process and the high uncertainty in the microfluidic cell.

\subsection{Validation of the relation between Soret coefficient and Gibb's free energy}
This section mainly focuses on validating Eq.\ref{eq:dG-T-high} at two different temperatures which connects  $\Delta G$ obtained from ITC with  $S_{\mathrm T}$ obtained from TDFRS. In the forthcoming sections, the calculation corresponds to T$_{\mathrm{high}} = 30^\circ $C and $T_{\mathrm{low}} = 20^\circ $C.
\subsubsection{EDTA-CaCl$_2$ system}
As mentioned before, the first system that we chose for the validation of the derived mathematical expression is EDTA-CaCl$_2$. With the $S_{\mathrm T}$ values of EDTA, CaCl$_2$ and the complex measured at T$_{\mathrm{high}}$ and T$_{\mathrm{low}}$, we have access to the change in Gibb's energy ($\Delta \Delta G$) of the individual components.  On the basis of our observations, we calculate  $\Delta G$ (30$^\circ$C) to be -36.5 $\pm$ 1.2 kJ mol$^{-1}$ , whereas from ITC measurements we obtained -36.4 $\pm$ 0.8 kJ mol$^{-1}$. Both values agree within the error limits. Repeating the calculations for other temperature pairs lead also to an agreement within 10\% (cf. Supporting Information).

\subsubsection{Protein-ligand system}
Now we apply the same procedure to the protein-ligand systems. In   Table \ref{delta_G} we compare the calculated $\Delta G$  and the measured $\Delta G_{\mathrm{ITC}}$. For both the ligands values agree well within the error bars. Values for other temperature pairs can be found in the Supporting Information.

\begin{table}[h!] 
	\	\caption{Comparison of $\Delta G$ that has been calculated and that has been measured with ITC\label{delta_G}}
	\newcolumntype{C}{>{\centering\arraybackslash}X}
	\begin{tabularx}{\textwidth}{CCCCC}
		\toprule
		System	& T$_{\mathrm{high}}$ ($^\circ$C) & T$_{\mathrm{low}}$ ($^\circ$C) & $\Delta$G$_\mathrm{calculated}$ (kJ/mol) & $\Delta$G$_\mathrm{ITC}$ (kJ/mol)\\
		\midrule
		BCA I+ PFBS		& 30		& 20 & -40.5 $\pm$ 1.1& -40.4$\pm$ 1.3\\
		BCA I + 4FBS		& 30			& 20 & -39.9 $\pm$ 3.9  & -38.2 $\pm$ 1.5\\
		
		\bottomrule
	\end{tabularx}
\end{table}

%%%%%%%%%%%%%%%%%%%%%%%%%%%%%%%%%%%%%%%%%%
\section{Materials and Methods}

\subsection{Sample preparation}

\subsubsection{EDTA-CaCl$_2$ system}
\label{Sec:EDTA_CaCl2_TDFRS_conc}
\noindent Stock solutions of EDTA and CaCl$_2$ were prepared in 2-(N-morpholino)ethanesulfonic acid (MES) buffer of 10 mM, pH 5.8. EDTA solution of 1 mM and CaCl$_2$ of 10 mM were used for measurements. For TDFRS samples, these solutions were filtered (0.2 $\mu$m) to remove dust particles. The transparent solution was filled into  an optical quartz cell (Hellma) with an optical path length of 0.2~mm. For ITC measurements, a calibration kit (Malvern Panalytical ) was used as received.

\subsubsection{BCA-ligand system}
\label{Sec:concentration_protein_ligand}
To prepare the ligand and protein solutions, sodium phosphate buffer (NaP buffer, pH 7.4, 20 mM) was used. Concentration of BCA I and ligand solutions were determined using UV-Vis absorption spectroscopy. Calibration curves (absorbance vs concentration) for BCA I, PFBS and 4FBS were prepared starting from the stock solution of 1 mg/ml and measuring the absorbance maxima at 280, 268 and 257 nm, respectively. For BCA I, the concentration of the solution was reconfirmed using molar extinction coefficient of BCA I (51.0 $\times$ 10$^3$ M$^{-1}$ cm$^{-1}$) and absorbance measured at 280 nm \cite{Osborne-1984-302}. For TDFRS experiments BCA I and ligand concentrations of 10 $\mu$M and 110 $\mu$M were used. For ITC experiments, the same concentration was used for BCA I + PFBS system, while for BCA I + 4FBS we had to increase protein and ligand concentrations to 20 $\mu$M and 300 $\mu$M respectively.

\subsection{Methods}

\subsubsection{Thermal Diffusion Forced Rayleigh Scattering}
Thermodiffusion of all the systems was measured by infrared thermal diffusion forced Rayleigh scattering (IR-TDFRS) \cite{Wiegand-2002-189, Blanco-2011-1602}. This method uses the interference grating of two infrared laser beams ($\lambda = 980$~nm) to generate a temperature grating inside an aqueous sample due to the inherent absorbtion of water at 980 nm \cite{Wiegand-2007-14169}. A third laser beam ($\lambda = 633$~nm) is refracted by this grating and the intensity of the refracted beam is measured. The intensity is proportional to the refractive index contrast of the grating, showing a fast rise over time due to the thermal gradient, then a slower change of intensity due to diffusion of the solute along the temperature gradient. The Soret, thermal diffusion and diffusion coefficient can be determined from the measurement signal when the refractive index contrast factors $(\partial n/\partial c)_{p,T}$ and $(\partial n/\partial T)_{p,c}$ are known \cite{Wiegand-2002-189}.

\subsubsection{Contrast factor measurement}
The change of refractive index with concentration $({\partial n/\partial c})_{p,T}$ was measured by a refractometer (Abbemat MW Anton Paar) at a wavelength of 632.8nm. Refractive indices for five concentrations at six different temperatures (20-45$^{\circ}$C) were measured to determine $({\partial n/\partial c})_{p,T}$.The concentration dependence of $n$ was linearly fitted to derive the slope $({\partial n/\partial c})_{p,T}$ for all measured temperatures. The refractive index increments with temperature $({\partial n/\partial T})_{p,c}$ were measured interferometrically \cite{Becker-1995-600}. Measurements were performed over a temperature range of 20-45$^\circ$C, with a heating rate of 1.6 mK/sec.

\subsubsection{Isothermal Titration Calorimetry}
\label{Sec:ITC_conc}
This technique has been extensively used to measure the thermodynamic parameters associated with protein-ligand binding interactions \cite{Velazquez-Campoy-2004-35}. When a ligand binds to a protein, thermodynamic potentials ($\Delta G$, $\Delta H$, $\Delta S$) change which can be measured by highly sensitive calorimetry. All other conventional methods measures binding affinity where as ITC measures the enthalpic and entropic contributions to binding affinity. This technique uses step wise injection of one reagent into the calorimetric cell. The working principle of the instrument has been discussed in the literature \cite{Velazquez-Campoy-2004-35,Baranauskiene-2009-2752,Thomson-2004-35}.

The calorimetric experiments for our study were performed with a MicroCal PEAQ ITC (Malvern Panalytical). For experiments on the reference system, EDTA (0.1 mM) in MES buffer (pH 5.8, 10 mM) was titrated with CaCl$_2$ (10 mM) in the same buffer at 6 different temperatures (20-45$^{\circ}$C with 5$^{\circ}$C gap). A typical experiment consisted of 19 injections, 2 $\mu$L each. The time interval between injections was 2.5 minutes. Measurements were conducted 2 times with a new stock solution of EDTA and CaCl$_2$ received from Malvern Panalytical. The same protocol was followed for BCA I-ligand sytems with concentrations as mentioned in Sec.\ref{Sec:concentration_protein_ligand}. For protein-ligand systems, measurements were also recorded at 6 temperatures between 20 and 45$^{\circ}$C at pH 7.4. Additionally, to study the effect of pH and labeling, extra measurements were carried out for BCA I-PFBS system. Binding of this system was monitored at 25$^{\circ}$C for two scenarios: (a)BCA I-PFBS at pH 8.3 and (b)labeled BCA I-PFBS at pH 8.3.
Data were analyzed using a single-site binding model subtracting background enthalpies, whereas  $\Delta H$ and $K_{d}$ are treated as adjustable parameters.

\subsubsection{Thermophoretic microfluidic cells}
The thermophoretic microfluidic cell can be either operated with large colloids (>500 nm), which are visible under the microscope or with fluorescently labeled macromolecules. The cell was made of PMMA and consisted of three channels \cite{Lee-2022-123002}. We created a 1D temperature gradient in the measurement channel between the heating and cooling channels. In order to measure the temperature and concentration profile in the channel, a confocal microscope (Olympus IX-71 with a FV3-294 confocal unit). A pulsed laser ($\lambda = 485$~nm) was used for probing the fluorescence intensity and lifetime. The fluorescence intensity for the concentration of proteins was measured by a photomultiplier and the fluorescence lifetime  in the measurement channel was characterized by fluorescence lifetime imaging microscopy (FLIM) using a correlator and a photomultiplier. The sample concentration of protein (BCA I) in the solution was 20 $\mu$M. The labeled protein content was 2.2 $\mu$M, which corresponds to 11\% of proteins in the solution.

%%%%%%%%%%%%%%%%%%%%%%%%%%%%%%%%%%%%%%%%%%
\section{Conclusions}
The main goal of this work is to investigate whether it is possible to connect thermodynamic parameters obtained by ITC with thermodiffusion parameters determined by TDFRS. For a low molecular weight reference system EDTA-CaCl$_2$ and the protein BCA I with two ligands 4FBS and PFBS we were able to relate Soret coefficients with the Gibb's free energies measured at two different temperatures with ITC using an empirical expression suggested by Eastman \cite{Eastman-1928-283}. This implies that Soret coefficients measured at different temperatures can be used to predict the Gibb's free energy at other temperatures. A second goal was to compare the results of the thermophoretic behavior of the protein and the complex with those obtained in a recently developed thermophoretic microfluidic cell. Fluorescent labeling of the protein is required to monitor the protein concentration using the thermophoretic microfluidic cell. For the here investigated protein BCA I, the fluorescent labeling influences the binding interactions severely so that we refrained from systematic thermophoretic measurements of the complex in the thermophoretic microfluidic cell. This is done more efficiently with an intrinsic fluorescent protein e.g. green fluorescent protein (GFP) or lysozyme. To gain a deeper microscopic understanding of the process, it would be desirable to perform neutron scattering experiments to determine the entropic contributions of the protein, thus unraveling the entire process\cite{Stadler-2015-72,Stadler-2016-72}.  

Further we found, that the Soret coefficients of EDTA and the EDTA+CaCl$_2$ complex show an unusual temperature dependence that cannot be described by Eq.\ref{eq_Piazza}. Of particular note is the abrupt drop in the Soret coefficient of EDTA between 25 and 30$^\circ$C. One finds some studies in the literature that also indicate a change in the behavior of EDTA in the same temperature range, but the data base is insufficient to draw clear conclusions. At this point more systematic pH-dependent measurements also of other chelating agents such as diethylene triamine pentaacetic acid (DTPA) or hydroxyethylethylenediaminetriacetic acid (HEDTA) would be desirable.

\begin{acknowledgments}
Most grateful we are to Olessya Yukhnoevets, who taught us the fluorescent labeling of the protein. We thank Mona Sarter and Andreas Stadler for fruitful and helpful discussions. We are grateful to Peter Lang and Jan Dhont for inspiring ideas and their generous support of our work. SM acknowledges the support by the International Helmholtz Research School of Bio-physics and Soft Matter (BioSoft) and NL acknowledges the support by the Humboldt foundation. 
\end{acknowledgments}

%\bibliographystyle{unsrt}
%\bibliography{protein-ligand}

\clearpage
\begin{widetext}
{
{\centering{\huge SUPPORTING INFORMATION: \\Complementary experimental methods to obtain thermodynamic 		parameters of protein ligand systems\par}
	
	{\Large Shilpa Mohanakumar, Namkyu Lee \\ and Simone Wiegand\par}
	
	{\large \today}\par}

\smallbreak

\section{Mathematical relation between Soret coefficient and Gibb's free energy}
\label{sec:math-G-ST}
Soret coefficient and Gibb's free energy at different temperatures have been measured by TDFRS and ITC, respectively. A relation between $S_{\mathrm{T}}$ starts from a relation originally proposed by Eastman \cite{Eastman-1928-283} and later rewritten by W{\"u}rger \cite{Wurger-2013-438} in a modern nomenclature,
\begin{equation}
	\label{eq:eastmanSI}
	S_{\mathrm T} =\frac{1}{k_{\mathrm B}T} \frac{dG}{dT}
\end{equation}
The Soret coeficients $S_{\mathrm T}^{\mathrm {low}}$ and $S_{\mathrm T}^{\mathrm {high}}$ correspond to $T_{\mathrm {low}}$ and $T_{\mathrm {high}}$, respectively.  Assuming a linear $T$-dependence of $S_{\mathrm T}$ with $T$, we write
\begin{equation}
	\label{Eq:linear_ST}
	S_{\mathrm T}(T)=(S_{\mathrm T}^{\mathrm {low}}+\Delta S_{\mathrm T})T
\end{equation}
with 
\begin{equation}
	\label{a_equation}
	\Delta S_{\mathrm T}={S_{\mathrm T}^{\mathrm {high}}-S_{\mathrm T}^{\mathrm {low}}}
\end{equation}
Integration of Eq.~\ref{eq:eastmanSI} with respect to temperature in the range from $T_{\mathrm{low}}$ to $T_{\mathrm{high}}$ leads to
\begin{equation}
	\label{ST_dG}
	\int_{T_{\mathrm {low}}}^{T_{\mathrm {high}}} dG =  \int_{T_{\mathrm {low}}}^{T_{\mathrm {high}}} k_{\mathrm B} T S_{\mathrm T}(T) dT
\end{equation}
using Eq.~\ref{Eq:linear_ST} we obtain
\begin{equation}
	\label{ST_dG_relation}
	\Delta \Delta G=  k_{\mathrm B}(S_{\mathrm T}^{\mathrm {low}}+\Delta S_{\mathrm T})(\frac{T_\mathrm {high}^{2}-T_\mathrm {low}^{2}}{2})
\end{equation}
For a mole of molecules holds 
\begin{equation}
	\label{ST_dG_relation_1mole}
	\Delta \Delta G= N_{\mathrm A} k_{\mathrm B}(S_{\mathrm T}^{\mathrm {low}}+\Delta S_{\mathrm T})(\frac{T_\mathrm {high}^{2}-T_\mathrm {low}^{2}}{2}).
\end{equation}
$S_{\mathrm T}$ values of ligand, macromolecule and the complex at two different temperatures, $T_{\mathrm {high}}$ and $T_{\mathrm {low}}$ can be measured with TDFRS. $\Delta \Delta G$ corresponding to ligand, macromolecule and the complex can then be calculated with  Eq.~(\ref{ST_dG_relation_1mole}). For systems like EDTA-CaCl$_2$ and protein-ligand for which binding is not strong, there are chances to find free ligands and macromolecules in the solution. In such cases, $S_{\mathrm T}(\mathrm {complex})$ will have also contributions from the individual compounds that are remaining in the solution. Hence there will be two main contributions to the $S_{\mathrm T}$ measured;
\begin{enumerate}
	\item Contribution from the complex 
	\item Contribution from free macromolecule or free ligand
\end{enumerate}
$S_{\mathrm T}(\mathrm {complex})$ for such systems has to be calculated excluding the contribution from free protein/free ligand (whichever is present in the system). It has to be noted that in EDTA-CaCl$_2$ system, EDTA acts as the macromolecule (in the cell of ITC) and CaCl$_2$ as the ligand (injected through  a syringe into the cell of ITC). Dissociation constant K$_{\mathrm d}$ measured by ITC is;
\begin{equation}
	\label{Dissociation_constant}
	K_\mathrm d=\frac{\mathrm{[protein]}\mathrm{[ligand]}}{\mathrm{[complex]}}
\end{equation}
where $\mathrm{[protein]}$, $\mathrm{[ligand]}$ and $\mathrm{[complex]}$ are the concentrations of free protein, free ligand and complex,  respectively. Since the total concentration of protein ($\mathrm {protein(total)}$) and ligand ($\mathrm {ligand(total)}$) is known, Eq.~(\ref{Dissociation_constant}) can be rearranged as follows, 
\begin{equation}
	\label{Dissociation_constant_rearranged}
	K_\mathrm d=\frac{[\mathrm {protein(total)}-[\mathrm {complex}]][\mathrm {ligand(total)}-[\mathrm {complex}]]}{[\mathrm {complex}]}
\end{equation}
Solving the quadratic equation Eq.~(\ref{Dissociation_constant_rearranged}) gives access to $[\mathrm {complex}]$, from which we can calculate the fraction of the free protein $f_\mathrm P$ and ligand $f_\mathrm L$ accordingly,
\begin{equation}
	\label{fraction_protein}
	f_\mathrm P=\frac{[\mathrm {protein(total)}-[\mathrm {complex}]]}{\mathrm {protein(total)}}
\end{equation}
\begin{equation}
	\label{fraction_ligand}
	f_\mathrm L=\frac{[\mathrm {ligand(total)}-[\mathrm {complex}]]}{\mathrm {ligand(total)}}
\end{equation}
\begin{equation}
	\label{fraction_complex}
	f_\mathrm C=1-(f_\mathrm P+f_\mathrm L)
\end{equation}
Analysis of K$_{\mathrm d}$ values indicates that for EDTA-CaCl$_2$ and both protein-ligand systems, the fraction of free ligand (or CaCl$_2$) is higher than that of the free protein (or EDTA) (at 25$^{\circ }$C, for BCA I-4FBS system, fraction of free protein and ligand are 0.008 and 0.7, respectively). Hence for the further calculations, we neglect the contribution from the free protein. We propose the following equation for the $S_{\mathrm T}\mathrm {(total)}$ that is measured using our IR-TDFRS setup  : 
\begin{equation}
	\label{ST_total}
	\frac{1}{S_{\mathrm T}^\mathrm {total}}=\frac{1-f_\mathrm C}{S_{\mathrm T}^\mathrm {complex}}+\frac{f_\mathrm L}{S_{\mathrm T}^\mathrm {ligand}}
\end{equation}
where $S_{\mathrm T}^\mathrm {total}$ is the $S_{\mathrm T}$ measured using IR-TDFRS, $S_{\mathrm T}^\mathrm {ligand}$ is the $S_{\mathrm T}$ of ligand, $f_\mathrm C$ is the fraction of complex and $f_\mathrm L$ is the fraction of free ligand. Rearranging Eq.~(\ref{ST_total}) leads to 
\begin{equation}
	\label{ST_complex}
	S_{\mathrm T}^\mathrm {complex}=\frac{1-f_\mathrm C}{\frac{1}{S_{\mathrm T}^\mathrm {total}}-\frac{f_\mathrm L}{S_{\mathrm T}^\mathrm {ligand}}}
\end{equation}
Access of $S_{\mathrm T}\mathrm {(complex)}$ at $T_\mathrm {high}$ and $T_\mathrm {low}$ enables the calculation of $\Delta \Delta G$ corresponding to the complex. Obtained values can be used to establish a relation between ITC and TDFRS measurements as shown in the main manuscript. 

\section{Various existing forms of EDTA in MES buffer}
\label{Sec:edta_forms}
EDTA (C$_{10}$H$_{16}$N$_2$O$_8$) exists in several forms in MES buffer \cite{Reiter-2015-thesis}. EDTA is a hexadentate ligand which has six lone pairs of electrons that are available for binding. Protonation enables EDTA to exist in various states ranging from a fully protonated (Fig.\ref{fig:forms_of_edta}a) to fully deprotonated stage (Fig.\ref{fig:forms_of_edta}b). It can also exist in some intermediate stages between these two forms with one to three protons. Experimental studies confirm that EDTA$^{4-}$ and HEDTA$^{3-}$ exist, playing a crucial role in forming complexes with the metal ion \cite{Kula-1966-697,Carr-1973-315,Reiter-2015-thesis}. It has to be noted that the complexation reaction primarily proceeds through EDTA$^{4-}$ forms, but HEDTA$^{3-}$ reacts also with a lower probablity \cite{Reiter-2015-thesis}. The majority form depends on the pH (cf. Fig.1 \cite{Kula-1966-697,Carr-1973-315} and Fig.9 in \cite{Reiter-2015-thesis}).
\begin{figure}[h!]
	\centering
	\begin{minipage}{.5\textwidth}
		\centering
		\includegraphics[width=7 cm]{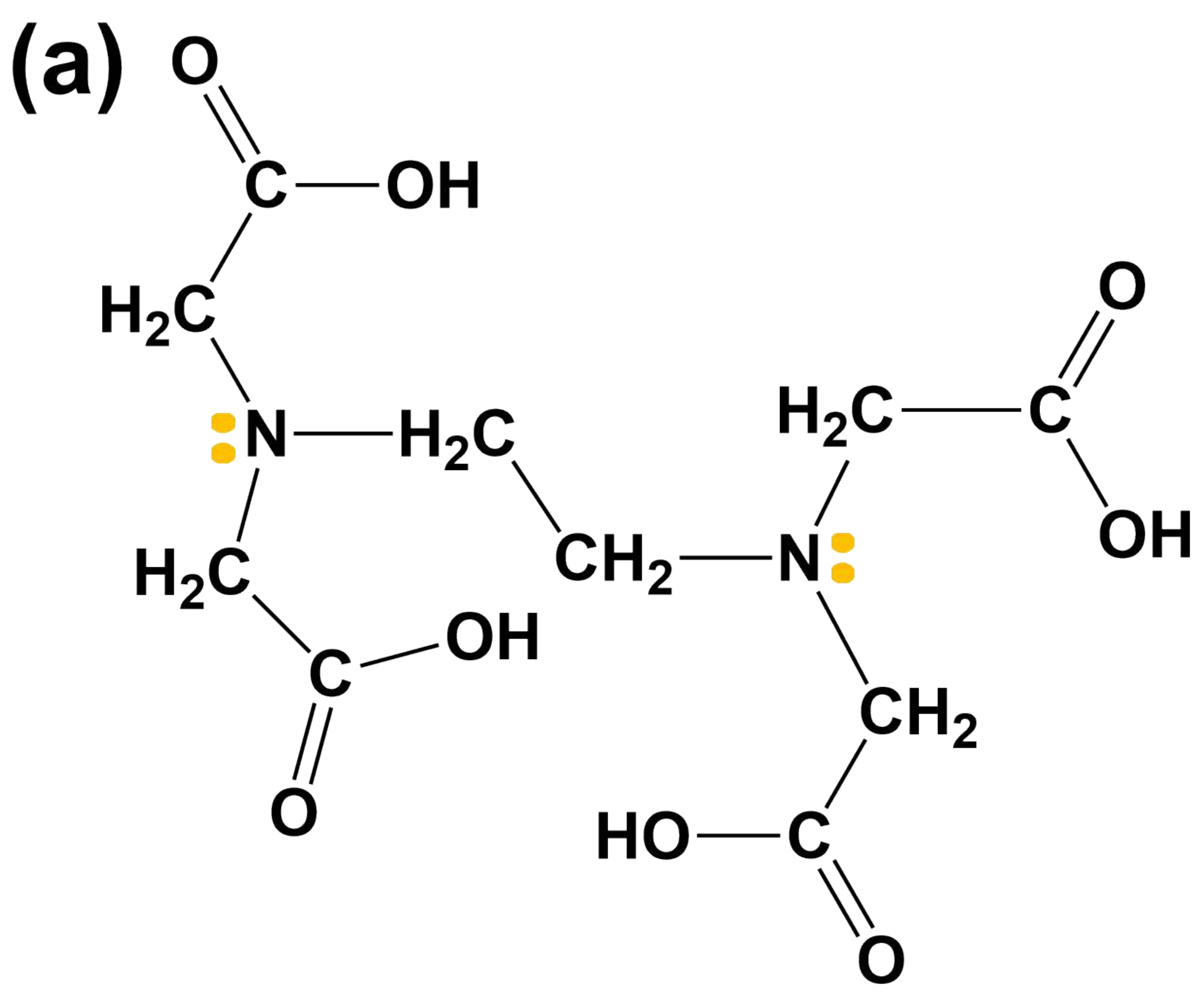}
		\label{fig:test1}
	\end{minipage}%
	\begin{minipage}{.6\textwidth}
		\centering
		\includegraphics[width=7 cm]{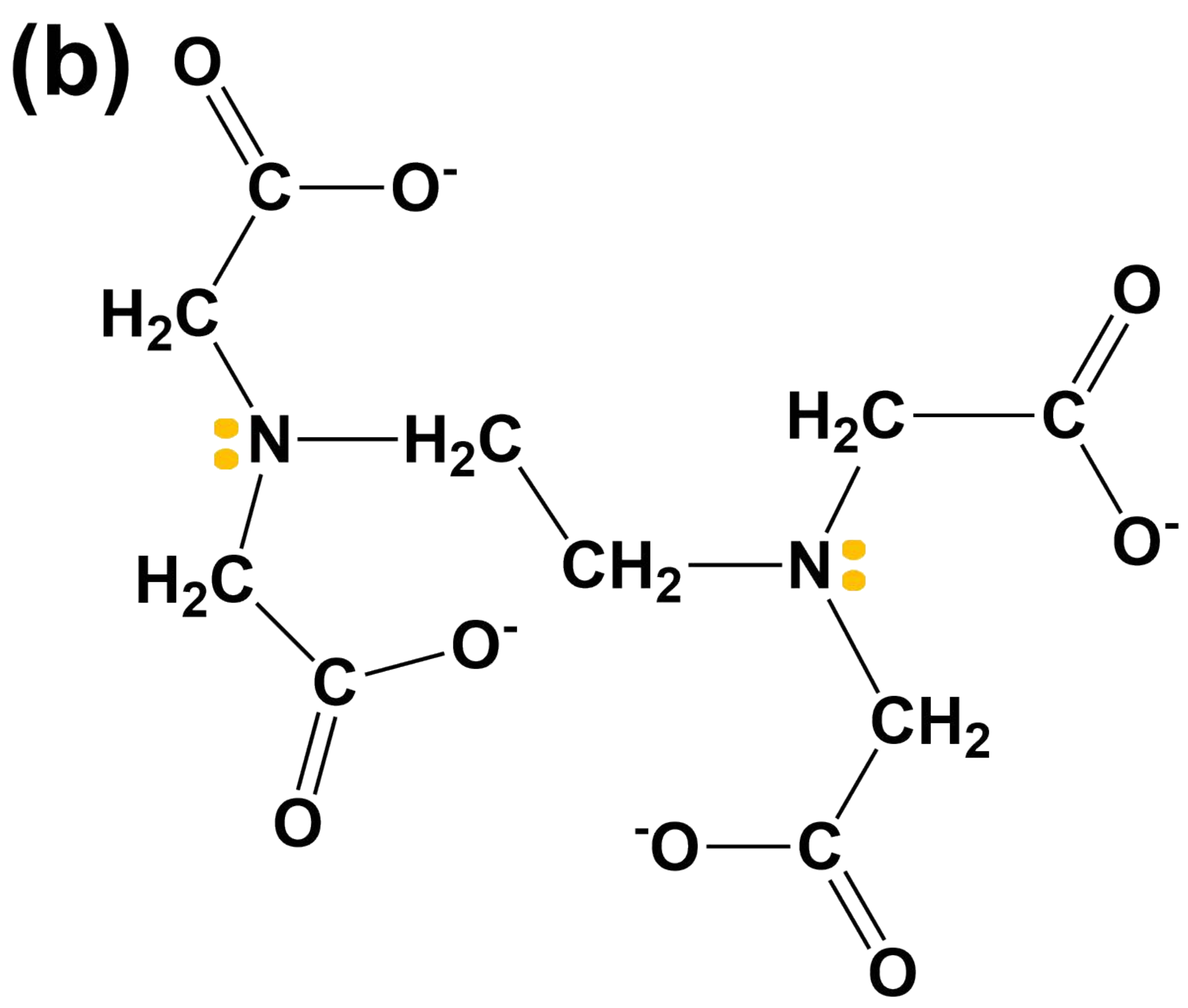}
		\label{fig:test2}
	\end{minipage}
	\caption{\label{fig:forms_of_edta}(a)Protonated form of EDTA (b)Deprotonated form of EDTA}
\end{figure}

\section{Protein-ATTO 532 dye labeling}
\label{Sec:labeling}

\subsection{Labeling BCA}
ATTO 532 dye (cf. Fig.\ref{fig:ATTO_532}) was used for labeling Bovine Carbonic Anhydrase (BCA) I. Protein standard (100 $\mu$M) was prepared in sodium phosphate buffer (20 mM). For NHS-ester labeling the pH of the sample was adjusted to 8.3 by adding drops of a NaHCO$_3$ solution (2.1 g in 50 ml water). 1 mg of ATTO 532 was dissolved in 200 $\mu$l of dimethyl sulfoxide (DMSO) which is used as the standard solvent for the dye. The dye solution was mixed with the protein solution so that the final concentration ratio dye:protein is 200~ $\mu$M:20~$\mu$M in 1 ml of the solution. The rest of the volume was filled up with buffer solution of pH 8.3. The sample were kept in shaker overnight for thorough mixing.
\begin{figure}[h!]
	\centering
	\includegraphics[width=8 cm]{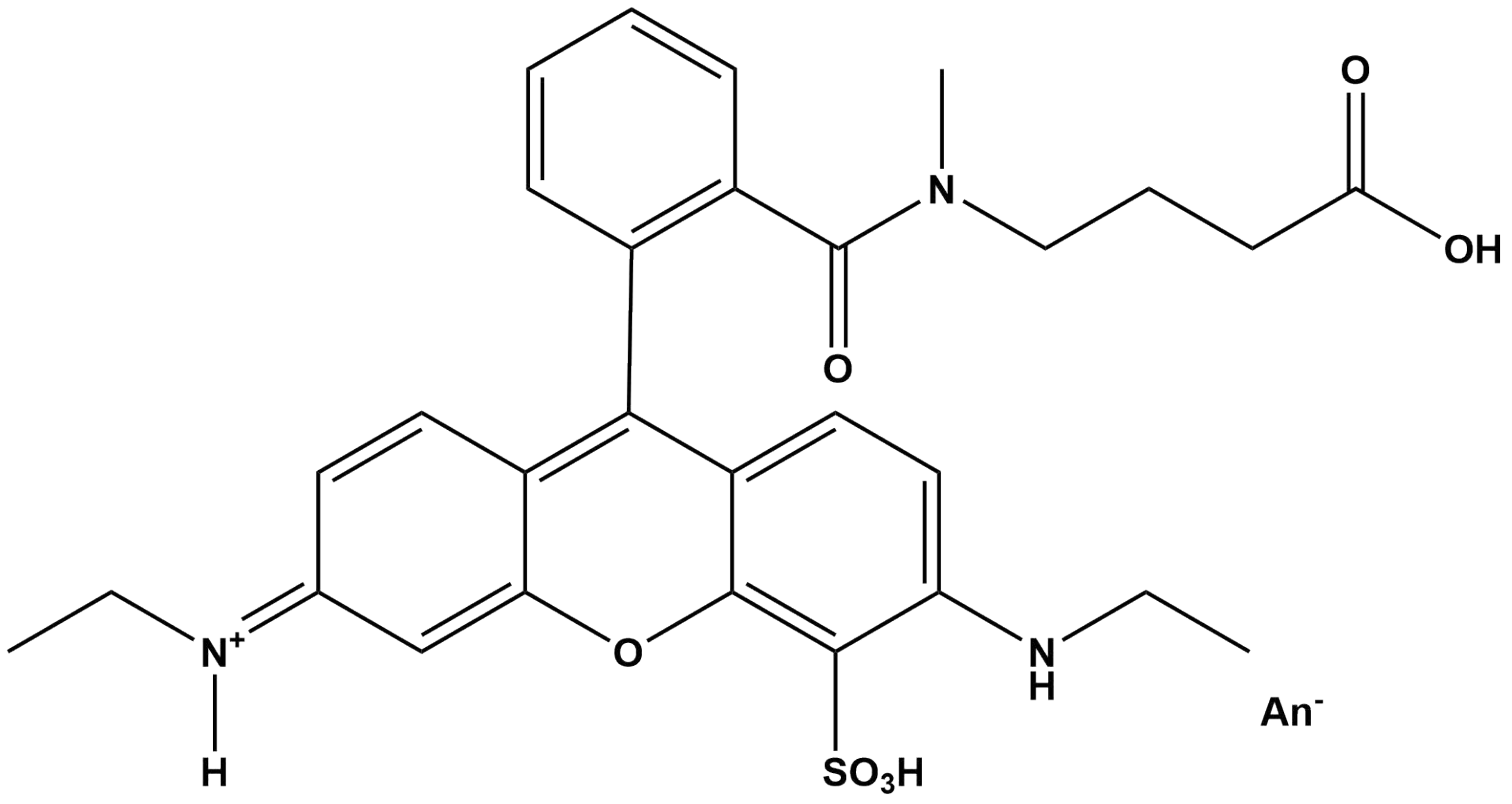}
	\caption{\label{fig:ATTO_532} Chemical structure of ATTO 532 dye}
\end{figure}
\subsection{BCA purification}
The following procedure was followed to remove the unbound dye.
Zeba™ Spin-Säulen columns were used for the purification of the mixture. Once the column is settled after centrifugation (1000 rpm for 2 minutes), the process was repeated with buffer of pH 7.4. 300 $\mu$l of the mixture was transferred to the column and purified. Absorption of the solution was measured using UV-VIS spectrophotometer. Labeled proteins have an absorption at 532 nm, which gives the corresponding concentration. 
\clearpage

\section{Temperature dependence of the thermal diffusion $D_{\mathrm{T}}$ and diffusion coefficient $D$ for protein-ligand systems}
\label{Sec:Temp_dependence}

\begin{figure}[h!]
	\centering
	\includegraphics[width=8 cm]{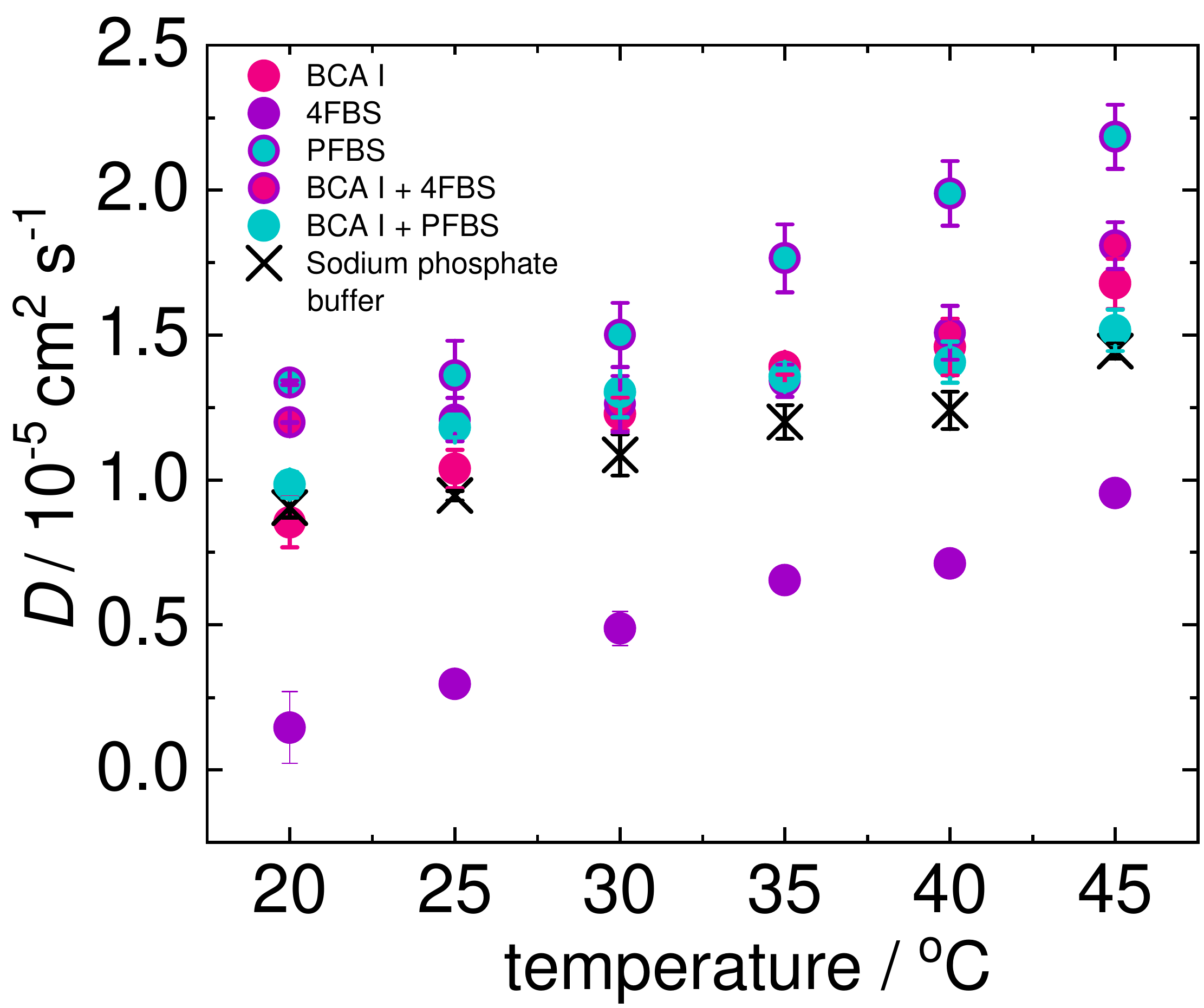}
	\caption{\label{fig:D_vs_T_BCA_ligand} Temperature dependence of $D$ for BCA I (protein), 4FBS and PFBS (ligands), corresponding protein-ligand complex, sodium phosphate buffer}
\end{figure}

Dependence of diffusion coefficients on temperature is shown in Fig.\ref{fig:D_vs_T_BCA_ligand}. Diffusion of protein gets faster when ligand binds to it. This could be related to the fact, that the complex is less hydrophilic, so that it can move faster compared to the free protein, which shows the slowest diffusion. The increase in $D$ with temperature due to decrease in viscosity can also be seen in Fig.\ref{fig:D_vs_T_BCA_ligand}. Dependence of thermodiffusion coefficient on temperature is shown in Fig.\ref{fig:DT_vs_T_BCA_ligand}. here it is striking that the thermal diffusion coefficients of the complexes and the ligands are very similar, which $D_{\mathrm{T}}$ is much lower, which could also be related to the higher hydrophilicity of the free protein.

\begin{figure}[h!]
	\centering
	\includegraphics[width=8 cm]{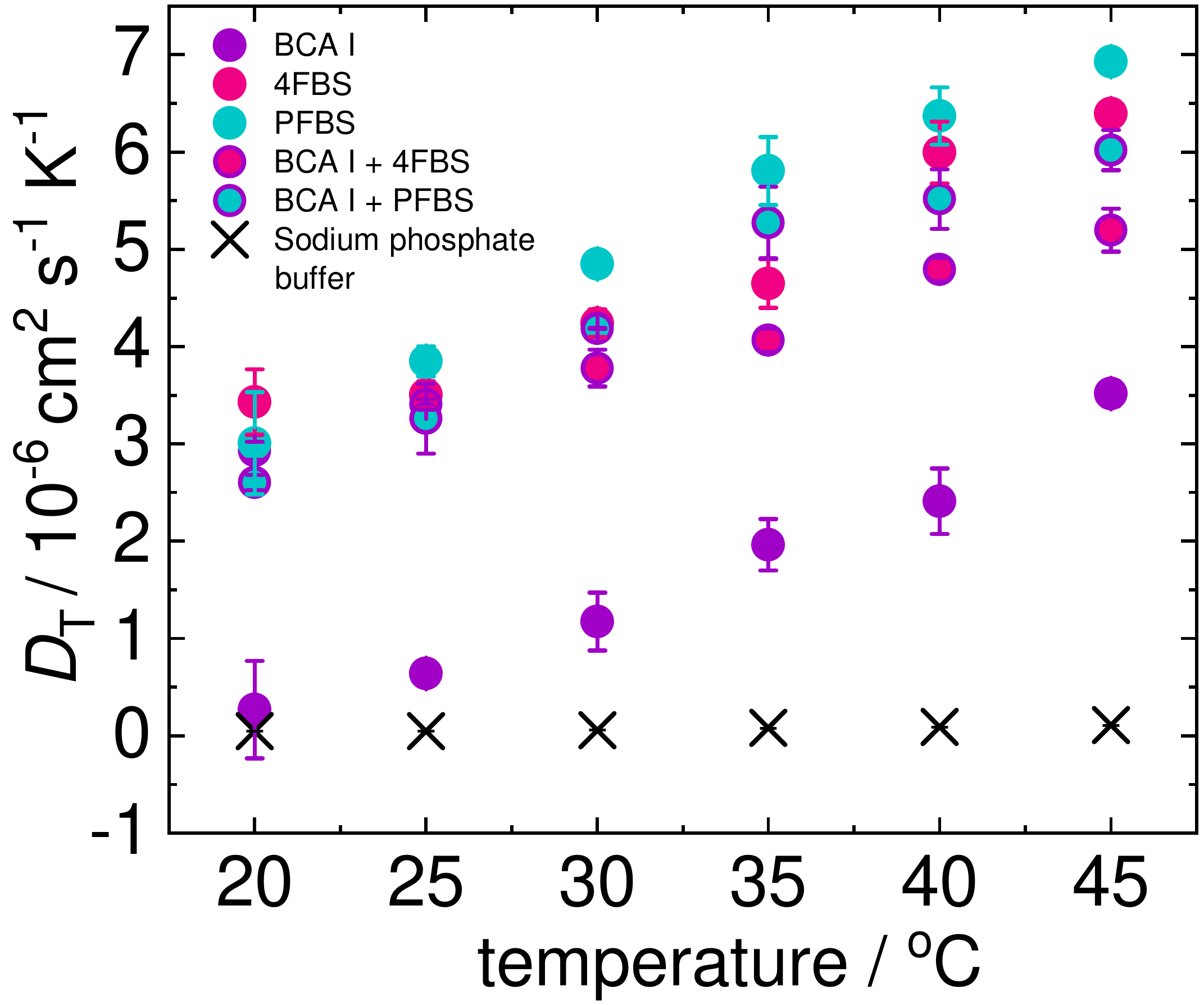}
	\caption{\label{fig:DT_vs_T_BCA_ligand} Temperature dependence of $D_{\mathrm{T}}$ for BCA I (protein), 4FBS and PFBS (ligands), corresponding protein-ligand complex, sodium phosphate buffer}
\end{figure}
\clearpage

\section{Refractive index increments with temperature}

Figure \ref{fig:dndc_dndt_edta_cacl2} and \ref{fig:dndc_dndt_protein_ligand} shows the refractive index increments for EDTA-CaCl$_2$ and both protein-ligand systems.

\begin{figure}[h!]
	\centering
	\includegraphics[width=8 cm]{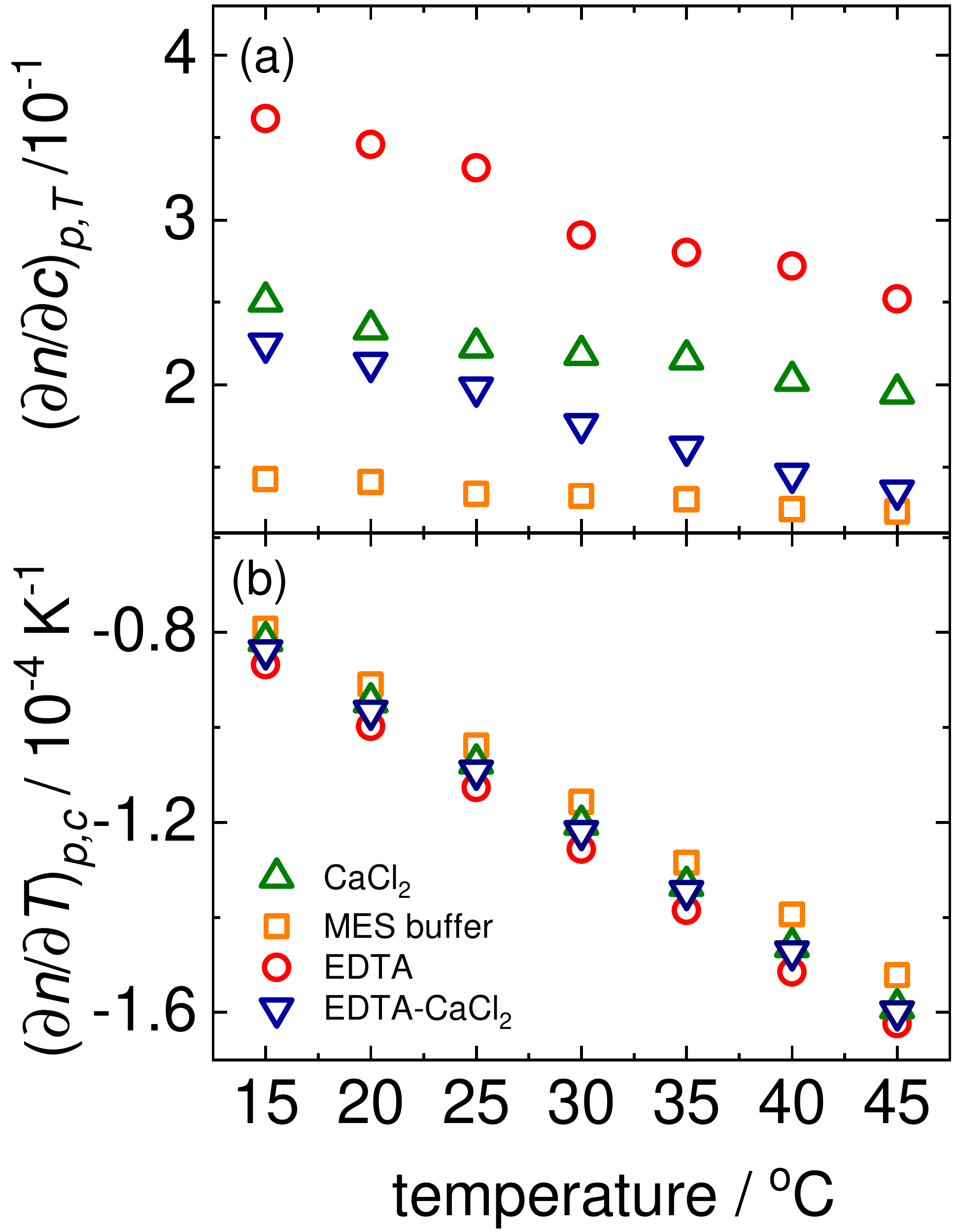}
	\caption{\label{fig:dndc_dndt_edta_cacl2}(a)Temperature dependence of $\left( \partial n / \partial c \right)_{p,T}$ for EDTA-CaCl$_2$ system (b)Temperature dependence of $\left( \partial n / \partial T \right)_{p,c}$ for EDTA-CaCl$_2$ system}
\end{figure}

\begin{figure}[h!]
	\centering
	\includegraphics[width=8 cm]{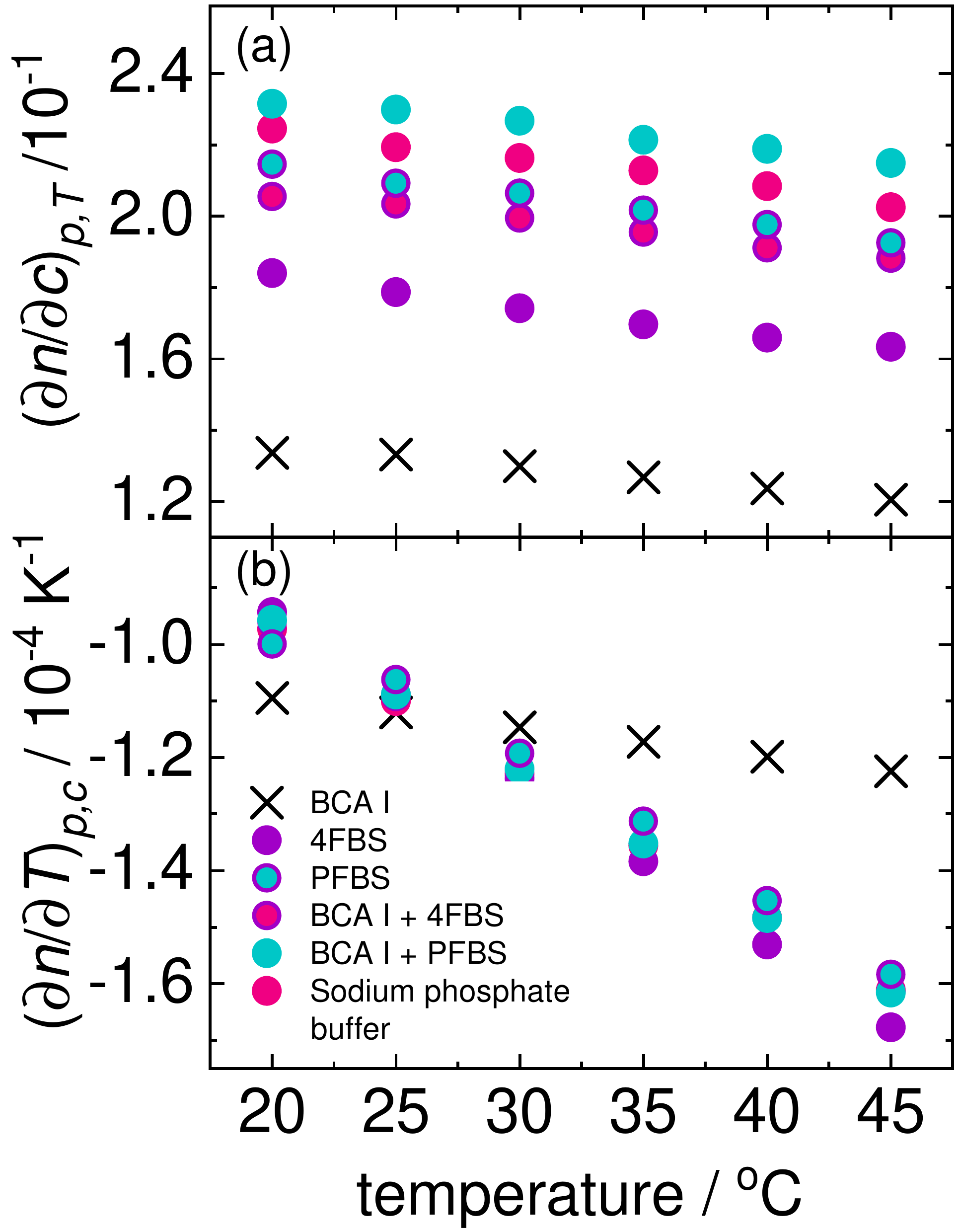}
	\caption{\label{fig:dndc_dndt_protein_ligand} (a)Temperature dependence of $\left( \partial n / \partial c \right)_{p,T}$ for protein-ligand systems (b)Temperature dependence of $\left( \partial n / \partial T \right)_{p,c}$ for protein-ligand systems}
\end{figure}
\clearpage

\section{Data analysis and fitting of ITC measurements}
%{\mynote{For me it is not clear which equation is used for fitting. Eq. (15) does not contain $n$ anymore.}}
We have used the most common binding model (1:1 binding) for the analysis of our ITC data. The following notations are used further: $m$ is the number of binding sites, $\theta$ is the fraction of sites occupied by ligand X, $M_t$ and $M$ are the bulk and free concentration of macro molecule, $X_t$ and $X$ are the bulk and free concentration of ligand, $\Delta H$ is the molar heat of ligand binding, $V_0$ is the active cell volume.
Binding constant,
\begin{equation}
	\label{eq:binding_constant}
	K=\frac{\theta}{(1-\theta)X}
\end{equation}
Total ligand concentration, 
\begin{equation}
	\label{eq:total_ligand}
	X_t=X+m\theta M_t
\end{equation}
Combining Eq.~\ref{eq:binding_constant} and Eq.~\ref{eq:total_ligand};
\begin{equation}
	\label{eq:theta}
	\theta^2-\theta[1+\frac{X_t}{mM_t}+\frac{1}{mKM_t}]+\frac{X_t}{mM_t}=0
\end{equation}
The total heat content of the solution $Q$ at a given time for an ITC measurement is 
\begin{equation}
	\label{eq:heat_of_solution}
	Q=m\theta M_t\Delta H V_0
\end{equation}
Solving Eq.~\ref{eq:theta} for $\theta$ and substituting it in Eq.~\ref{eq:heat_of_solution} gives, 
\begin{equation}
	\label{eq:heat_of_solution_new}
	Q=\frac{mM_t\Delta H V_0}{2}[1+\frac{X_t}{mM_t}+\frac{1}{mKM_t}- \sqrt{(1+\frac{X_t}{mM_t}+\frac{1}{mKM_t})^2-\frac{4X_t}{mM_t}} ]
\end{equation}
Once an injection of ligand into the cell occurs, software calculates the heat associated with the injection. What is of primary interest for the study is the change in heat content of $i^{th}$ injection to that of $i-1^{th}$ injection.
Change in heat after  $i^{th}$ injection is given by;
\begin{equation}
	\label{heat_release}
	\Delta Q(i)=Q(i)+\frac{dV_i}{V_0}[\frac{Q(i)+Q(i-1)}{2}]-Q(i-1)
\end{equation}
where $dV_i$ is the correction factor introduced to compensate the heat contribution from the displaced volume. This is because, for each injection there is a volume of protein-ligand complex that is expelled from the cell which is identical to the volume of ligand that is injected.  Fitting involves initial guesses of $m$, $\Delta H$ and $K$ which allows to calculate $\Delta Q(i)$ for each injection. This is then compared with the experimental $\Delta Q(i)$ which is measured and then improving $m$, $\Delta H$ and $K$ values until the best fit. A typical ITC measurement curve and molar enthalpy curve is shown in Fig.\ref{fig:ITC_measurement_BCAI_PFBS}.
\begin{figure}[h!]
	\centering
	\includegraphics[width=8 cm]{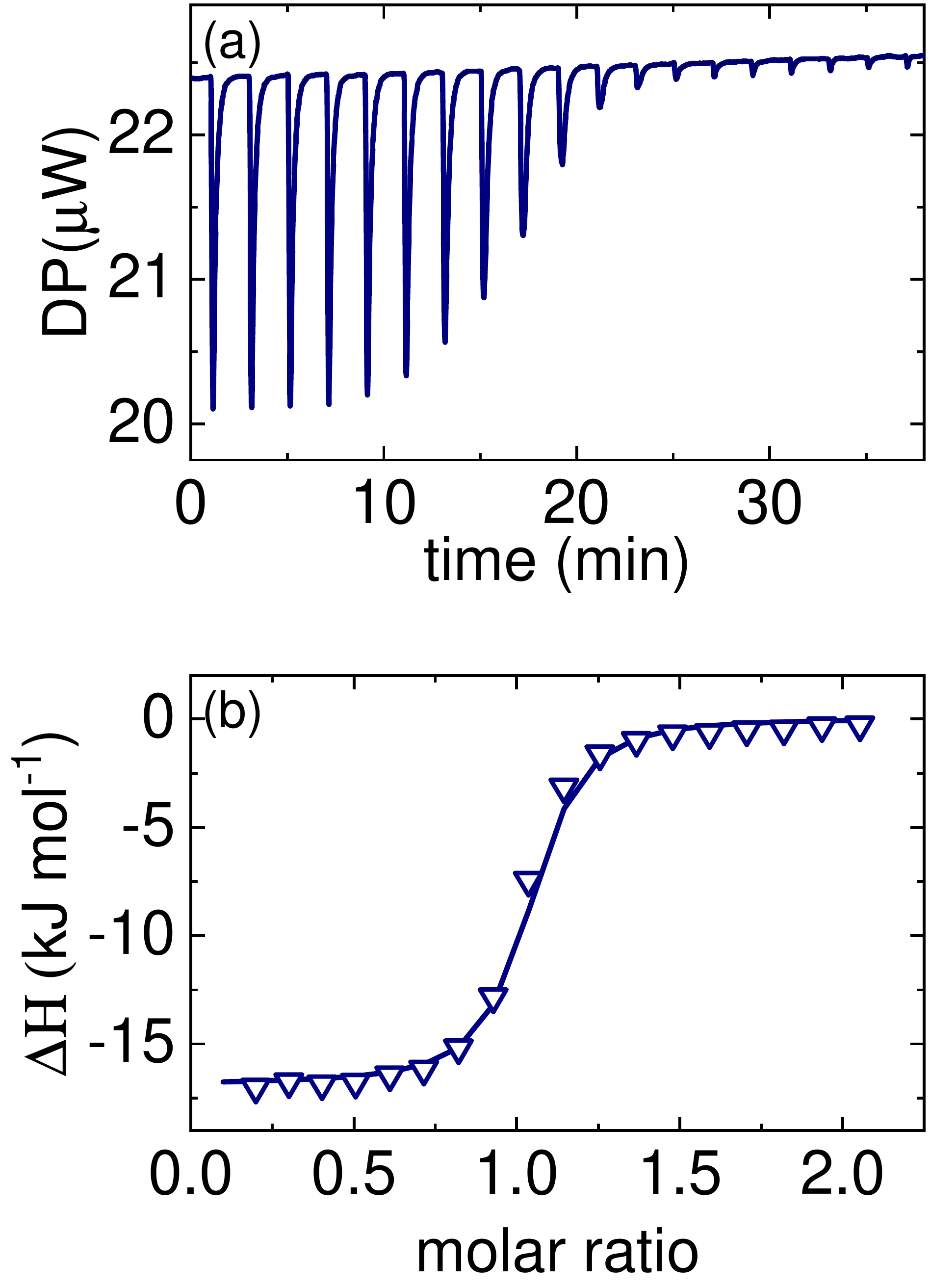}
	\caption{\label{fig:ITC_measurement_BCAI_PFBS} (a)Raw data output for EDTA-CaCl$_2$ binding reaction measured at 25$^\circ$C (b)Integrated data output for EDTA-CaCl$_2$ binding reaction measured at 25$^\circ$C plotted as molar change in enthalpy against molar ratio }
\end{figure}
\clearpage

\section{Validation of the relation between Soret coefficient and Gibb's free energy at other temperatures}
$\Delta$G calculated at higher temperatures using Eq.\ref{eq:dG-T-high} (cf. Sec.1 of main manuscript) and its comparison to the ITC measurements will be discussed for all systems in the following paragraphs. 
\subsection{EDTA + CaCl$_2$}
Table \ref{tb1:dG_calculation_edta_cacl2} lists the calculated $\Delta$G$_\mathrm{calculated}$ and $\Delta$G$_\mathrm{ITC}$ measured with ITC for EDTA + CaCl$_2$. Both values agree within the error bars. 
\begin{table} [h!]
	\caption{\label{tb1:dG_calculation_edta_cacl2} Table enlists $\Delta$G values that are calculated and measured using ITC at different temperature combinations for EDTA-CaCl$_2$ system}
	\begin{center}
		\begin{tabular}{|c|c|c|c|}
			\hline 
			T$_{\mathrm{high}}$ ($^\circ$C) & T$_{\mathrm{low}}$ ($^\circ$C) & $\Delta$G$_\mathrm{calculated}$(kJ/mol) & $\Delta$G$_\mathrm{ITC}$(kJ/mol) \\ \hline
			20 &  30      & -36.5 $\pm$ 1.2   & -36.4 $\pm$ 0.8     \\ \hline  
			25 &  35      & -36.1$\pm$ 3.2     & -35.5 $\pm$ 1.8     \\ \hline  
			30 &  40      & -35.7 $\pm$ 3.4     & -34.8 $\pm$ 2.4     \\ \hline 
			35 &  45      & -33.3 $\pm$ 1.4     & -34.2 $\pm$ 1.9   \\ \hline

		\end{tabular} 
	\end{center}
\end{table}

\subsection{Protein-ligand}
\subsubsection{BCA I + PFBS}
Table \ref{tb1:dG_calculation_bca_pfbs} lists the calculated $\Delta$G$_\mathrm{calculated}$ and $\Delta$G$_\mathrm{ITC}$ measured with ITC for BCA I + PFBS. Both values agree within the error bars. 
\begin{table} [h!]
	\caption{\label{tb1:dG_calculation_bca_pfbs} Table enlists $\Delta$G values that are calculated and measured using ITC at different temperature combinations for BCA I - PFBS system } 
	\begin{center}
		\begin{tabular}{|c|c|c|c|}
			\hline 
			T$_{\mathrm{high}}$ ($^\circ$C) & T$_{\mathrm{low}}$ ($^\circ$C) & $\Delta$G$_\mathrm{calculated}$(kJ/mol) & $\Delta$G$_\mathrm{ITC}$(kJ/mol) \\ \hline
			20 &  30      & -40.5 $\pm$ 1.1     & -40.4 $\pm$ 1.3   \\ \hline  
			25 &  35      & -44.0 $\pm$ 2.6     & -46.8 $\pm$ 0.6   \\ \hline  
			30 &  40      & -48.2 $\pm$ 3.1     & -52.1 $\pm$ 1.2    \\ \hline 
			35 &  45      & -54.9 $\pm$ 2.9     & -55.9 $\pm$ 1.1     \\ \hline

		\end{tabular} 
	\end{center}
\end{table}

\subsubsection{BCA I + 4FBS}
Table \ref{tb1:dG_calculation_bca_4fbs} lists the calculated $\Delta$G$_\mathrm{calculated}$ and $\Delta$G$_\mathrm{ITC}$ measured with ITC for BCA I + PFBS. Both values agree within the error bars. 
\begin{table} [h!]
	\caption{\label{tb1:dG_calculation_bca_4fbs} Table enlists $\Delta$G values that are calculated and measured using ITC at different temperature combinations for BCA I - 4FBS system } 
	\begin{center}
		\begin{tabular}{|c|c|c|c|}
			\hline 
			T$_{\mathrm{high}}$ ($^\circ$C) & T$_{\mathrm{low}}$ ($^\circ$C) & $\Delta$G$_\mathrm{calculated}$(kJ/mol) & $\Delta$G$_\mathrm{ITC}$(kJ/mol) \\ \hline
			20 &  30      & -39.9 $\pm$ 3.9    & -38.2 $\pm$ 1.5   \\ \hline
			25 &  35      & -36.6$\pm$ 3.4     & -39.4 $\pm$ 1.1   \\ \hline  
			30 &  40      & -40.3 $\pm$ 1.8     & -40.7 $\pm$ 0.8    \\ \hline 
			35 &  45      & -45.5 $\pm$ 3.5    & -42.1 $\pm$ 1.3     \\ \hline

		\end{tabular} 
	\end{center}
\end{table}

\clearpage

%\bibliographystyle{unsrt}
%\bibliography{protein-ligand}

}
\end{widetext}

\end{document}